\documentclass[12pt]{iopart}

\usepackage{hyperref}
\usepackage{graphicx}
\usepackage{rotating}
\usepackage{array}

\def\be{\begin{equation}}
\def\ee{\end{equation}}
\def\bea{\begin{eqnarray}}
\def\eea{\end{eqnarray}}

\def\ie{{\it i.e.\,}}
\def\etal{{\it et al.   }}

\def\<{\langle}
\def\>{\rangle}

\begin{document}

\title{Exact propagators on the lattice with applications to diffractive effects}
\author{E. Sadurn\'i }

\address{Instituto de F\'isica, Benem\'erita Universidad Aut\'onoma de Puebla,
Apartado Postal J-48, 72570 Puebla, M\'exico}

\eads{ \mailto{sadurni@ifuap.buap.mx}}

\begin{abstract}

The propagator of the discrete Schr\"odinger equation is computed and its properties are revealed through a Feynman path summation in discrete space. Initial data problems such as diffraction in discrete space and continuous time are studied analytically by the application of the new propagator. In the second part of this paper, the analogy between time propagation and 2D scattering by 1D obstacles is explored. New results are given in the context of diffraction by edges within a periodic medium. A connection with tight-binding arrays and photonic crystals is indicated.

\end{abstract}

%\submitto{\JPA}

\maketitle

\section{Introduction}

The formulation of Quantum Mechanics in discrete variables is relevant to various aspects of undulatory physics. The applications range from solid state (tight-binding arrays \cite{10}, polymers \cite{9}, photonic structures \cite{7}) to fundamental questions concerning the structure of space (discrete canonical variables \cite{hooft1}, cellular automata \cite{hooft2}, algebraic deformations and bound states \cite{20}). The dynamics of this kind of systems seems to be a natural area of study, filling a gap between the energy domain description and a time domain formulation yet to be explored. Moreover, among the many dynamical problems that one may pose, we distinguish diffraction by edges and the propagation of discontinuities (in space and in time \cite{2}) as the most prominent representatives. In continuous space, these examples contain the effects of infinitesimal wavelengths. In discrete space, the natural frequency cut-off might be used to conjecture new effects. But beyond these conjectures, it is necessary to solve the propagation problem of specific distributions. 

In this paper we tackle some time-dependent problems by computing explicitly the propagator of an infinite tight-binding chain. This is equivalent to the discretized one-dimensional space propagator or the Green's function of a two-dimensional photonic structure in the lowest frequency band. The mathematical tools employed here, although of a simple nature (Bessel functions), constitute an original contribution of this work. They shall be thouroughly developed in order to get new results in the problem of diffraction. This applies to propagation in a periodic background and the effects produced by a minimal spacing between points.

The paper is structured as follows: In section 2 we compute the propagator for the discrete Schr\"odinger equation by spectral decomposition (one or many dimensions for cubic lattices). As this object differs from its continuous counterpart, it is necessary to reveal its properties and its connection with Feynman paths in space and time. This shall be done in sections 2.1 and 2.2 respectively. In section 3 we move to specific time-dependent problems by applying the propagator to distributions of the edge type: A point-like distribution viewed as a minimal length wavepacket (section 3.1), the discrete version of the Moshinsky shutter (section 3.2) and a poissonian profile generalizing the former examples (section 3.3). Section 4 establishes the connection between diffraction in time in discrete space and the scattering of two-dimensional waves by a screen. Here it is shown that the scattering problem in a periodic background (such as that provided by a photonic structure) can be viewed as a problem of propagation along the optical axis, with periodicity corrections given by our newly obtained kernel (\ref{e7}). Section 5 gives a more detailed explanation of the relation between discrete space and periodic structures, casting the propagator in terms of continuous variables (\ref{e103}). We end with a brief summary.

\section{Discrete propagator in one dimension}

Let us start with the free Schr\"odinger problem in discrete space and continuous time. We proceed by discretizing the derivative operator in the continuous wave equation. The corresponding wave function can be written as $\phi_n(\tau)$ with $n$ an integer denoting a point in discrete space and $\tau$ a real parameter denoting time. Introducing the constants $\Delta$ (with units of energy $\times$ length$^2$) and $a$ (lattice spacing), the Schr\"odinger dynamical problem is described by the equation

\bea
-\frac{\Delta}{2 a^2} \left[ \phi_{n+1}(\tau) -2\phi_n(\tau) + \phi_{n-1}(\tau) \right] = i \hbar \frac{\partial \phi_n(\tau)}{\partial \tau}
\label{e1}
\eea
or, more concisely

\bea
-\frac{1}{2}\left[\psi_{n+1}(t) + \psi_{n-1}(t) \right] = i \frac{\partial \psi_n(t)}{\partial t},
\label{e2}
\eea
which is obtained under the appropriate rescaling of time $t=\frac{\Delta}{\hbar a^2} \tau$ and the transformation $\psi_n(t)= e^{it}\phi_n(\tau)$. The eigenvalue problem corresponding to (\ref{e2}) can be solved analitycally by means of Bloch waves \cite{bloch}:

\bea
-\frac{1}{2}\left[\psi^k_{n+1} + \psi^k_{n-1} \right] = E_k \psi^k_n,
\label{e3}
\eea

\bea
\psi^k_n = \frac{1}{\sqrt{2\pi}}e^{ikn}, \qquad E_k = -\cos k, \qquad k \in (0,2\pi)
\label{e4}
\eea
This simple set of solutions gives rise to a spectral decomposition of the evolution operator

\bea
U_{n,m}(t) = \int_0^{2\pi} dk \psi^k_n (\psi^k_m)^* e^{-iE_k t}
\label{e5}
\eea
and the solution of any initial-data problem satisfying (\ref{e2}) can be written as

\bea
\psi_n(t) = \sum_{m \in \bf{Z}} U_{n,m}(t)\psi_m (0).
\label{e6}
\eea
We are therefore in the position to write the propagator for positive times as $K(n,m;t)= \theta(t) U_{n,m}(t)$, where it is left to compute the integral in (\ref{e5}). We proceed by using the Jacobi-Anger expansion \cite{1} of $e^{-iE_k t}$ leading to Bessel functions of the first kind, namely 

\bea
K(n,m;t) &=&  \frac{1}{2\pi} \theta(t)  \int_0^{2\pi} dk e^{ik(n-m)} e^{i \cos k t} \nonumber \\
&=&  \frac{1}{2\pi}  \theta(t) \int_0^{2\pi} dk e^{ik(n-m)}\sum_{l \in \bf{Z}} i^{l} e^{ikl} J_l(t)  \nonumber \\
&=&  \theta(t)  i^{n-m} J_{n-m}(t),
\label{e7}
\eea
where we have used the Kronecker delta from the integration of the exponentials and the property $J_{-n}=(-1)^n J_n$ in the last step. Noteworthy is the existence of a phase, which shall prove to be of utmost importance in the analysis of Feynman paths in further sections. The meaning of such a phase can be readily given by expressing (\ref{e7}) in polar form: $K(n,m;t) = e^{i \pi L/2} J_{n-m}(t)$, where $L=n-m$ is a discrete length between endpoints.

By one stroke, we can give the propagator for $d$-dimensional lattices of the cubic type: Let $\bf{n}, \bf{m}$ be vectors denoting points in the lattice. The total propagator contains one-dimensional kernels appearing in the form of factors, leading to the overall result

\bea
K(\textbf{n}, \textbf{m}; t) = \theta(t) \prod_{j=1,...,d}  i^{n_j-m_j} J_{n_j-m_j}(t).
\label{e8}
\eea
It is important to note that the propagators of more complicated lattices with nearest-neighbour couplings cannot be cast as the simple product above, a problem that shall be studied in a follow-up paper \cite{5} for $d=2$. In the following we verify the properties of $K(n,m;t)$ by means of a few theorems on Bessel functions \cite{1}. 

\subsection{Properties of the new propagator}

The fundamental properties of continuous propagators can be verified in the discrete case as well. It is instructive to see that some of the well-known identities of Bessel functions have a meaning for our kernel. Recent studies on Bessel functions such as summation formulae \cite{1.1} and operational methods \cite{1.2} can be further utilized in the context of propagation. In this section we restrict ourselves to the most important properties. We finish by showing that the continuous limit of (\ref{e7}) leads to the possibility of computing corrections due to a minimal spacing $a$.
\begin{itemize}

\item Connection with the identity: It can be shown directly by evaluating the Bessel function using the ascending series. One has $K(n,m;0)=i^{n-m}J_{n-m}(0)=i^{n-m} \delta_{n-m,0}=\delta_{n,m}$.

\item Composition law: This shall be useful in a formulation of the propagator in terms of paths. The Neumann theorem for the addition of Bessel functions (see p. 358 \cite{1}) ensures this property. Let $t_1>0, t_2>0$. We obtain
\bea
\fl \sum_{l \in \textbf{Z}}K(n,l;t_1)K(l,m;t_2) &=& i^{n-m}\sum_{l \in \textbf{Z}} J_{n-l}(t_1) J_{l-m}(t_2) \nonumber \\ &=& i^{n-m} J_{n-m}(t_1+t_2)=K(n,m;t_1+t_2)
\label{e11}
\eea
\item The propagator as Green's function of the discrete Schr\"odinger equation: For $t>0$, $K(n,m;t)$ satisfies the homogeneous equation (\ref{e2}) by construction. To see this in terms of $J_n(z)$, we find that the recurrence relations for the derivatives of Bessel functions $2J'_{n}(z)=J_{n-1}(z)-J_{n+1}(z)$ (see formula (2), p. 17  in \cite{1}) yield the result directly. When $n=m$ and $t$ is allowed to vanish, we recognize that 

\bea
\fl \frac{1}{2}\left[ K(n+1,n;t) + K(n-1,n;t) \right] - i \frac{\partial K(n,n;t)}{\partial t} = -i \delta(t)K(n,n;t) = -i \delta(t),
\label{e12}
\eea
implying thus
\bea
\fl \frac{1}{2}\left[ K(n+1,m;t) + K(n-1,m;t) \right] - i \frac{\partial K(n,m;t)}{\partial t} = -i \delta(t) \delta_{n,m}.
\label{e13}
\eea
\item Continuous limit: Here we restore our units by introducing again the quantity $a$. When $a \rightarrow 0$, we keep $a(n-m)=\textrm{constant} \equiv x-x'$ as our continuous variables, with the obvious implication that $n-m$ must be large. In this limit, $t \sim \tau/a^2$ becomes large as well, but as a consequence $(n-m)/t = O(a) \rightarrow 0$. Therefore we invoke the asympotitc form of Bessel functions when both the argument and the index are large quantities (see C. VIII in \cite{1}). This is the so-called approximation by tangents \cite{3}, where the argument of $J$ is replaced by $\nu \sec \beta$ such that $\beta \rightarrow \pi/2$. The corrections are under control by recognizing that the approximation can be expressed as the first term of a series through Meissel's second formula (p. 227 in \cite{1}). We have 

\bea
 J_{n-m} \left(t\right) &\equiv& J_{\nu}(\nu \sec \beta) \approx \frac{\exp \left[ i \nu \left( \tan \beta - \beta \right) \right]}{\sqrt{2\pi \nu}(1-\sec^2 \beta)^{1/4}}
\label{e14}
\eea
and using $\sec \beta \equiv z$ with $z \gg 1$, the Taylor expansions in $1/z$ lead to

\bea
 J_{n-m} \left(t\right) & \approx & \frac{(-i)^{\nu}e^{i t} \exp \left[ i \frac{\nu^2}{2t} \right]}{\sqrt{2 \pi i t}} \nonumber \\
&=& (-i)^{n-m} e^{it} \sqrt{\frac{\hbar a^2}{2 \pi i \tau \Delta}} \exp \left[ i \frac{\hbar (x-x')^2}{2 \tau \Delta}  \right]
\label{e14.1}
\eea

Finally, we identify $\Delta = \hbar^2 / m$, where $m$ is the mass of a quantum particle. The gauge factor $e^{it}$ is due to our simplification in the discretization of the second derivative (the step leading to (\ref{e2}) from (\ref{e1})), whereas the factor of $a$ is the infinitesimal length to be absorbed as a differential when we consider the summation of amplitudes, \ie $\sum_{m} a \mapsto \int dx'$. The factor of $i^{m-n}$ is cancelled by the prefactor in the last line of (\ref{e7}). These steps lead to

\bea
K(n,m;t) \rightarrow \left[ a e^{it} \right] \times \sqrt{\frac{m}{2 \pi i \hbar \tau}} \exp \left( i  \frac{ m (x-x')^2}{2 \hbar \tau} \right)
\label{e15}
\eea
which corresponds to the free propagator in continuous variables. It is amusing to see that the next order correction of Meissel's expansion leads to the first corrections of the propagator due to discreteness in space (or minimal length). This shall be investigated in the context of diffraction.

\end{itemize}

\subsection{Connection with Feynman's sums over paths}

Now we investigate a formulation of discrete propagators leading to a summation over paths. The points to be touched upon are the existence of a phase associated to an action and a time-dependent weight function as in the usual Feynman path integral \cite{3.1}.

\begin{figure}[!h] \begin{center} \begin{tabular}{cc} \includegraphics[scale=0.6]{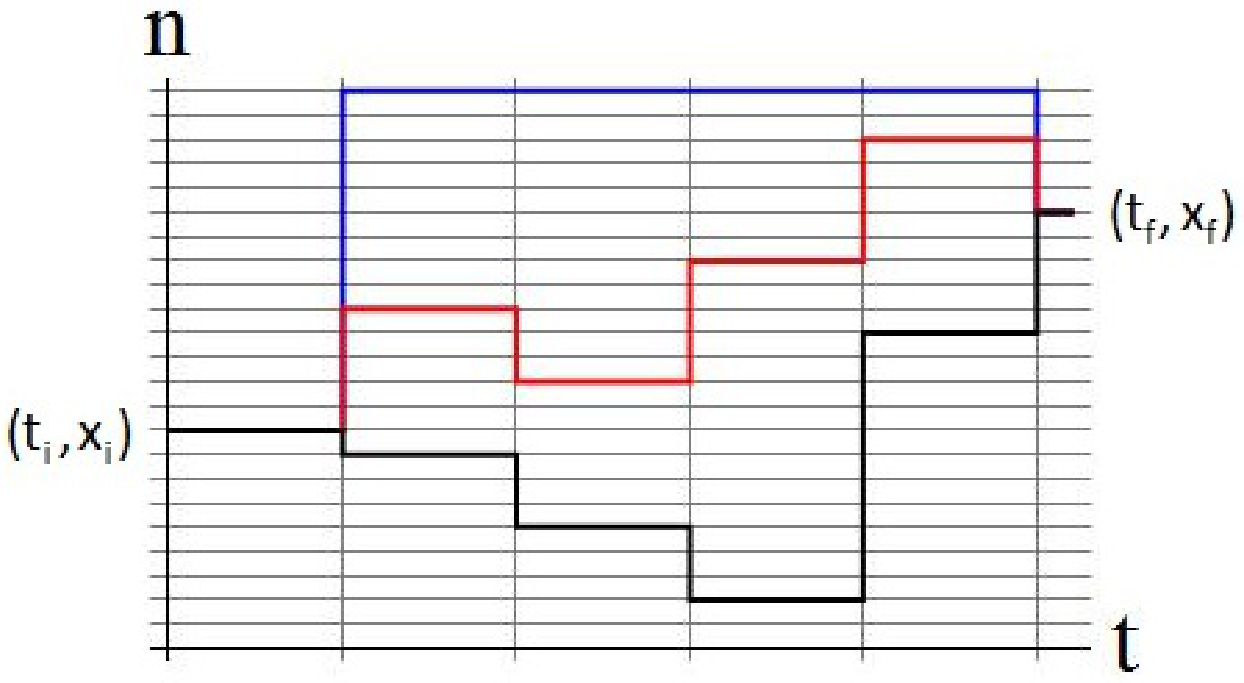} &  \includegraphics[scale=0.6]{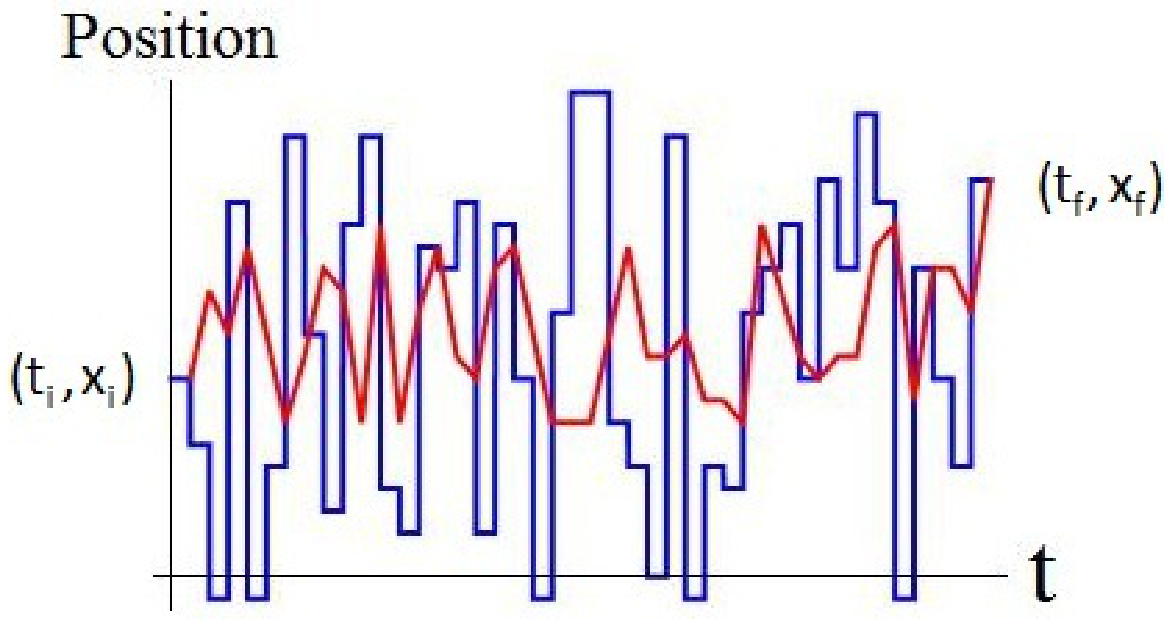} \end{tabular} \end{center} 

\caption{ \label{fig:1} Left panel: Three discontinuous paths of equal length joining $(t_i,x_i)-(t_f,x_f)$. The blue path contains only one change in direction. Right panel: Comparison of a typical continuous path in Feynman's integrals (Red curve) and a discontinuous trajectory in our path formulation (Blue curve)} \end{figure}

It is often pointed out that the original idea of composing propagators for small time intervals, comes from the fact that each propagator is a phase factor containing the classical action \cite{3.1, 3.2, 3.3}.  In discrete problems, however, we do not have such a classical action and Feynman's recipe cannot be applied here straightforwardly. We recognize that the formalism of path integrals was designed to deal with problems in continuous space \cite{note1}. Nevertheless we may employ
the composition property of discrete propagators (\ref{e11}) in order to investigate the process of multiple compositions along an arbitrary number of time slices. Denoting such a number as $N+1$ and defining $\nu_{N+1} \equiv n, \nu_{0} \equiv m, t_{N+1}\equiv t, t_0 \equiv 0$, we can express our propagator as the sum of products

\bea
K(n,m;t) &=& \sum_{\nu_1, \dots, \nu_N} \prod_{j=0}^N K(\nu_{j+1},\nu_{j};t_{j+1}-t_{j}) \nonumber \\
&=&  \sum_{\nu_1, \dots, \nu_N} \prod_{j=0}^N i^{\nu_{j+1}-\nu_{j}} J_{\nu_{j+1}-\nu_{j}}(t_{j+1}-t_{j})
\label{e21}
\eea
where the times are ordered as $t_{j+1}>t_{j}$ for each slice $j$. As a side remark, this expression is yet another representation of a Bessel function which, to the author's knowledge, has not been investigated in the mathematical literature. We proceed to do so, by taking equal time intervals $t_{j+1}-t_{j}= t / (N+1)$. We should note also that the kernel (\ref{e7}) is symmetric $K(n,m;t)=K(m,n;t)$ as a consequence of $J_{-n}(z)=(-)^n J_{n}(z)$ and of the inversion of the complex prefactor. Then, it is legitimate to replace $i^{n-m}J_{n-m}(t)=i^{|n-m|}J_{|n-m|}(t)$ for each factor in the r.h.s. of (\ref{e21}). Furthermore, we express the complex prefactors $i^{|n-m|}$ in polar form to obtain

\bea
K(n,m;t) &=& \sum_{\nu_1, \dots, \nu_N} \prod_{j=0}^N  e^{i \pi |\nu_{j+1}-\nu_{j}| / 2} J_{|\nu_{j+1}-\nu_{j}|}(t/(N+1))  \nonumber \\
&=&  \sum_{\nu_1, \dots, \nu_N} \exp \left( i \frac{\pi}{2} \sum_{j=0}^{N} |\nu_{j+1}-\nu_{j}|  \right) \prod_{j=0}^N J_{|\nu_{j+1}-\nu_{j}|}(t/(N+1)) \nonumber \\
&=&  \sum_{\nu_1, \dots, \nu_N} \exp \left( i \frac{\pi}{2} S_{N+1,0} \right) \prod_{j=0}^N J_{S_{j+1,j}}(t/(N+1))
\label{e22}
\eea 
where in the last line we have identified $S_{N+1,0}= \sum_{j=0}^{N} |\nu_{j+1}-\nu_{j}|$ as the {\it total length\ } of a path joining the events $(0,m)$ and $(t,n)$. Similarly for $S_{j+1,j}$. Therefore, the coherence of the sum is completely governed by the phase $S_{N+1,0}$, which we are tempted to call {\it action\ }:  The minimization of this functional gives a reduced class of paths containing, in particular, the discrete straight line. One might call such a trajectory a {\it classical path,\ }although our dynamical problem has been defined exclusively for discrete quantum systems. The sum is dominated by the paths satisfying $S_{min}=n-m$. We should note that the terms in the sum are also affected by a {\it weight\ } $W \equiv \prod_j J_{S_{j+1,j}}(t/N)$. Despite of being real, such a weight dictates the contribution of each term and is, at first sight, a path-dependent function. This is evident from the fact that each Bessel function contains explicitly the partial length $S_{j+1,j}$ in its index. In order to understand better this sum, we study the weights in the limit $N \rightarrow \infty$. Taking the first term in the ascending series of the Bessel function (formula (8) p. 40 in \cite{1}), and denoting each path $\nu_1,...,\nu_n = \{ \nu\}$, we can make explicit the separation between path-independent contributions $w(t)$ and path-dependent contributions $F[\{ \nu \}]$:

\bea
W = \left(\frac{t}{2(N+1)}\right)^{S_{N+1,0}} \times \prod_{j=0}^{N} \frac{1}{(S_{j+1,j})!} \equiv  w(t) \times F[\{\nu \} ].
\label{e23}
\eea
The final sum reads

\bea
K(n,m;t) = \sum_{ \textrm{\scriptsize Paths} } w(t) F[ \{ \nu \}] \exp \left( i \frac{\pi}{2} S_{N+1,0} \right)  \nonumber \\
\label{e24}
\eea 
and with this expression we find that Feynman's prescription applies to discrete problems with the appropriate amendments, \ie  a factor $F$ correcting the weights. The partial path lengths are the key, as we have shown. The presence of $F$ can also be analyzed on a conceptual basis: The paths contributing to our propagator are not in the same class as those appearing in continuous-variable path integrals. The latter turn out to be continuous curves, the majority of them being nowhere differentiable. In contrast, the paths that we are studying are discontinuous by assumption, jumping in $\textbf{Z}$ (depicted in figure \ref{fig:1}). The paths that might approach dangerously to a non-measurable function in space-time are killed (or supressed) by the weight $F$. The paths that are favoured by $F$ are those whose lengths do not approach to infinity and contain a finite number of discontinuities. And this is possible only if the continuous pieces of such curves are constant functions of $t$. This is a clear generalization of the integrals appearing in diffusive processes involving the Wiener measure \cite{3.3}, as well as a deviation from the traditional path integral formulation. It is, however, consistent with our mathematical procedures.

To complete the circle, we finish this section by showing that taking (\ref{e24}) as a starting point, we arrive again at a single Bessel function. First, we perform a change of summation index: Let $\{ S \} = S_{N+1,N},...,S_{1,0}$ be the set of partial lengths characterizing a path $\{ \nu \}$ of a fixed length $S$. We can separate the sum into

\bea
\sum_{ \{ \nu \}  } = \sum_{S=S_{min}}^{\infty} C(S) \times \sum_{ \{ S \} }
\label{e25}
\eea 
where $C(S)$ is a coefficient counting the number of paths of equal length. Obtaining $C(S)$ is a combinatorial task and the procedure can be found in Appendix A. We have

\bea
C(S)= \frac{S!}{(\frac{S+n-m}{2})!(\frac{S+m-n}{2})!}
\label{e26}
\eea
and noting that $S_{min}=n-m$, the sum (\ref{e24}) becomes

\bea
\fl K(n,m;t) =  \sum_{S=n-m}^{\infty}\frac{S!}{(\frac{S+n-m}{2})!(\frac{S+m-n}{2})!}  \exp \left( i \frac{\pi}{2} S \right)  \sum_{ \{ S \} } \left(\frac{t}{2(N+1)}\right)^{S}  \prod_{j=0}^{N} \frac{1}{(S_{j+1,1})!}.
\label{e27}
\eea 
Moreover, we observe that the factor $F$ is related to a multinomial coefficient, leading to

\bea
\fl \sum_{ \{ S \}: \textrm{\scriptsize length}=S } \left(\frac{t}{2(N+1)}\right)^{S} \frac{S!}{  \prod_{j=0}^{N} (S_{j+1,1})!} = \left( \sum_{j=0}^{N}\frac{t}{2(N+1)} \right)^{S} = \left(\frac{t}{2} \right)^{S}.
\label{e28}
\eea
Finally, the propagator can be written as an ascending series in $t$

\bea
\fl K(n,m;t) &=&  \sum_{S=n-m}^{\infty} \frac{ \exp \left( i \frac{\pi}{2} S \right)}{(\frac{S+n-m}{2})!(\frac{S+m-n}{2})!}  \left(\frac{t}{2} \right)^{S} \nonumber \\
&=& i^{n-m}\sum_{s=0}^{\infty}   \frac{ (-)^{s}}{(s+n-m)!s!}  \left(\frac{t}{2} \right)^{2s+n-m} \nonumber \\
&=& i^{n-m} J_{n-m}(t), \qquad t>0.
\label{e29}
\eea
We have shown that discrete quantum mechanics, in its simplest form (the free problem), allows a description in terms of paths and weights. The resulting paths are discontinuous at a finite number of points in time and generalize the paths of diffusion processes. Moreover, the prescription we have obtained can be used again to recover a closed formula for the propagator $-$ A book-keeping combinatorial procedure.

\section{Applications and new effects}

An important question about discrete and continuous quantum systems is related to the effects emerging in the presence of a minimal length. The obvious example in which we can distinguish clear differences comes in the form of diffraction, where it is known \cite{2, 4} that in the case of continuous variables, an infinite number of wavelengths is necessary in order to describe the propagation of edges (for example, diffraction in time). A direct consquence of space discreteness is the irrelevance of the Gibbs phenomenon in the Fourier decomposition of square packets. However we do not contempt ourselves with this simple conclusion: We extend further our study by describing the propagation of initial conditions such as the discrete version of the Moshinsky shutter \cite{2} and the point-like initial condition.

\subsection{The expansion of a one-cell packet}

We start by analyzing the propagation of a packet of minimal width: one cell or {\it pixel.\ }In the continuous regime, such an initial condition cannot be conceived, but it is well-known that a packet with a square shape \cite{11, 4} gives rise to non-trivial oscillations before the natural regime of expansion occurs. Let $\psi(0)$ be given by

\bea
\psi_n (0) = \delta_{n,0}.
\label{e31}
\eea
Contrary to the case of a point-like stimulus in continuous space, our evolved solution yields a wavefunction with non-trivial density in the position variable $n$:

\bea
\psi_n(t)= i^{n} J_{n}(t), \qquad |\psi_n(t)|^2 =  J^2_{n}(t).
\label{e32}
\eea
 \begin{figure}[!h] \begin{center} \includegraphics[scale=0.7]{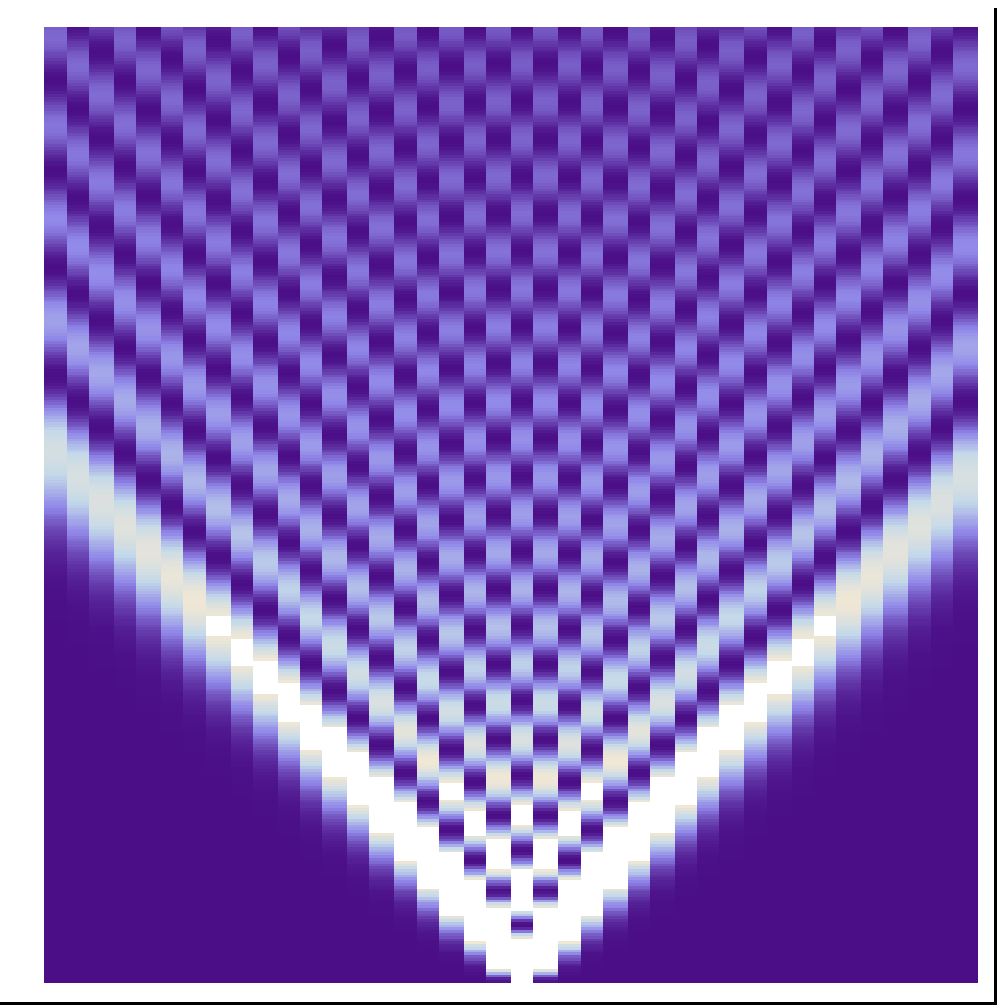}  \includegraphics[scale=0.7]{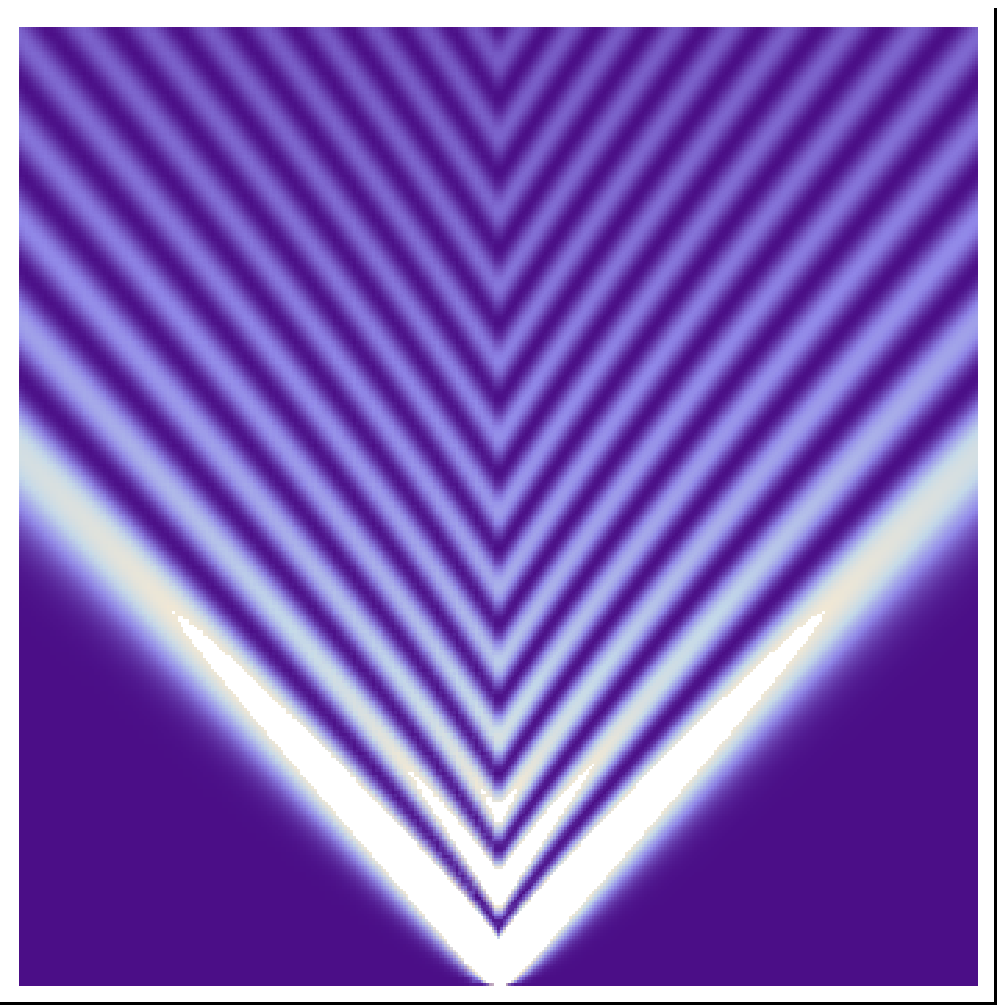} \end{center} 

\caption{ \label{fig:2} Left panel: Probability density in the plane $n$ (abscissa) and $t$ (ordinate) of a point-like source. We can see an expansion of the density at a constant velocity (set as unity). The expansion is accompanied by oscillations between the fronts $n\pm t$. These features cannot be found in the propagation of a point-like source in continuous variables. Right panel: Increased resolution, emulating a continuous variable $n$ in the abscissa.} \end{figure}

We can see the behaviour of the probability density in the plane $(n,t)$ in figure \ref{fig:2}. As the most prominent feature of this evolution, we find a regime of expansion which extends for all times. This is in sharp contrast with the continuous-variable problem, where infinitesimal oscillations typically appear around discontinuities in the initial conditions \cite{11, 4}. The edges are known to produce patterns approximately describable by the Cornu spiral, but the geometrical explanation provided by such a curve is not applicable to our discrete example. In fact, the regime of oscillations around an edge (the analog of Fresnel diffraction) is absent: We only find an expanding pattern with oscillations {\it inside\ }two wavefronts (bright diagonal fringes). The absence of the Fresnel regime is not the only notable effect. The natural frequency cut-off provided by the discreteness of the problem ($E_{max} = \Delta/a^2$) suggests a maximal velocity of propagation, which can be taken as $\Delta/\hbar a$ or unity in our conventions. The bright fringes of unit slope give rise to a conical structure in space and time, despite of the non-relativistic nature of the problem. The bright fringes indicate an abrupt change in the density, but not an actual discontinuity of the density function in the $t$ variable $-$ They appear due to a valley typically found in Bessel functions of large index. This peculiar slope or velocity shall appear again in the problem of the Moshinsky shutter.

We note, in passing, that the aformentioned wavefronts can be related to caustics in the Bessel integral. These caustics can be easily determined by finding the stationary point with the phase function $k(n-m)+t \cos k$. An appropriate description of the function around the caustics comes from the expansion of the Bessel function in terms of Airy functions \cite{nist}, covering the regions inside and outside of the conical structure (the Meissel expansions mentioned before cannot cover both regions simultaneously). 

\subsection{The Moshinsky shutter: Diffraction in discrete space and continuous time}

Now we consider, for negative times, a free wave on the left semi-axis moving to the right. A blocking (absorptive) screen at the position $n=0$ prevents the wave to propagate to the positive axis. In order to mimic Moshinsky's problem, such a wave must be taken as a Bloch wave, which solves the problem without boundaries. At $t=0$ the shutter is removed and the evolution studied:

\bea
\psi_n(0) = \theta(-n)e^{ikn}, \qquad \theta(0) \equiv 1
\label{e33}
\eea
\bea
\psi_n(t) &=& \sum_{m=-\infty}^{0} i^{n-m}J_{n-m}(t) e^{ikm} \nonumber \\
&=& i^{n}\sum^{\infty}_{m=0}  J_{n+m}(t) e^{-i(k-\pi/2) m} \nonumber \\
& \equiv &  i^{n}\sum^{\infty}_{m=0}  J_{n+m}(t) \xi^{m}
\label{e34}
\eea
Although the series above has the form of a generating function with complex parameter $\xi$, there is no closed form reported in the literature for this expansion (the summation index does not take negative values). However, we can extract some properties of this solution by exploiting some identities. For example, we can find the value of $\psi$ at position $n$ when the wave function at the shutter position is known, \ie

\bea
\psi_{n}(t) = e^{ikn} \left[ \psi_0(t) - \sum_{m=0}^{n-1}   J_{m}(t) e^{-i(k-\pi/2) m} \right]
\label{e35}
\eea
where the limit of the sum becomes {\it finite.\ }The function $\psi_0(t)$ has notable properties, for instance

\bea
\psi_{0}(t)|_{k=q-\pi/2} + \psi_{0}(t)|_{k=\pi/2-q} =  e^{it \sin q} + J_0 (t) . 
\label{e36}
\eea
A quick proof of this formula can be obtained by direct substitution of the second line of (\ref{e34}) on the l.h.s. of (\ref{e36}). Then by means of the Jacobi-Anger expansion we have $e^{it \sin q}= J_0(t)+\sum_{n=0}^{\infty} J_n(t) \left(e^{inq}+(-)^n e^{-inq}\right)$  and the r.h.s. of (\ref{e36}) follows.

\begin{figure}[!h] \begin{center} \begin{tabular}{ccc} \includegraphics[scale=0.5]{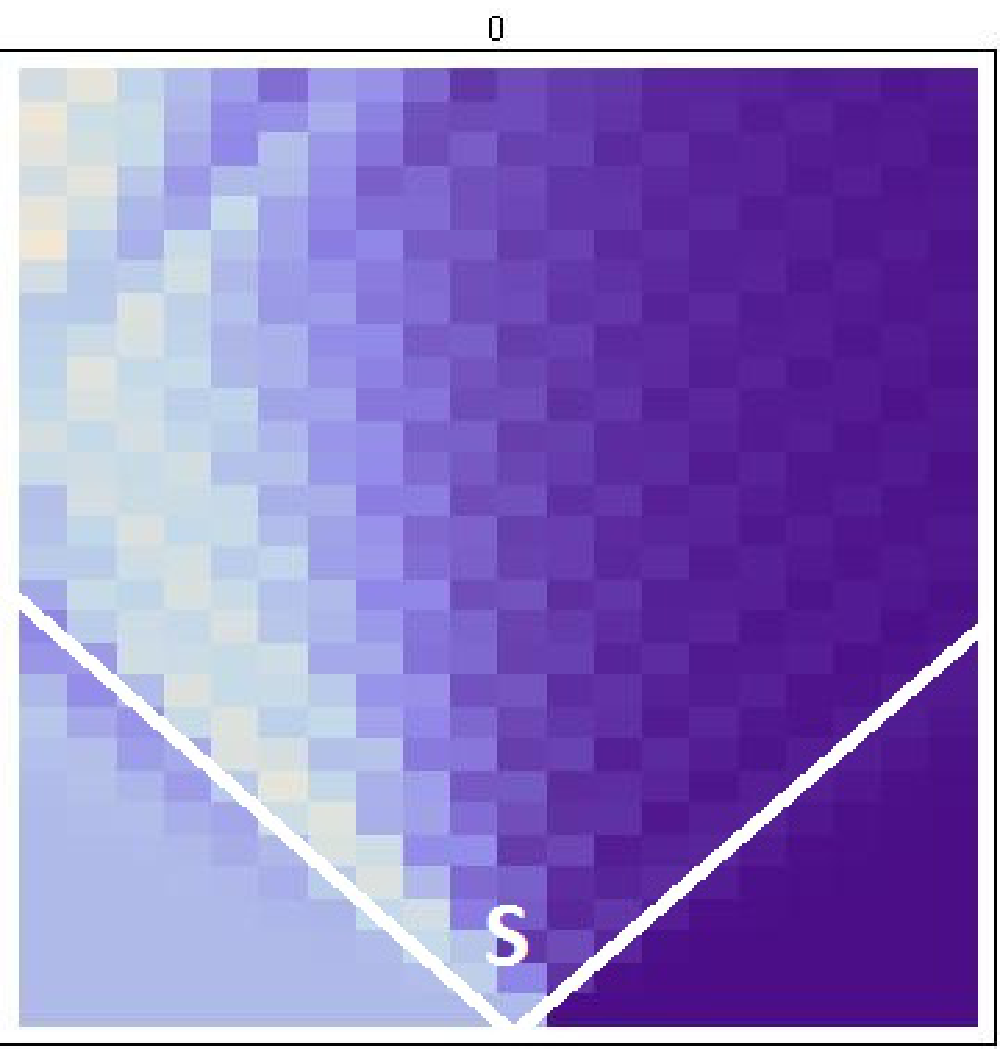} &  \includegraphics[scale=0.5]{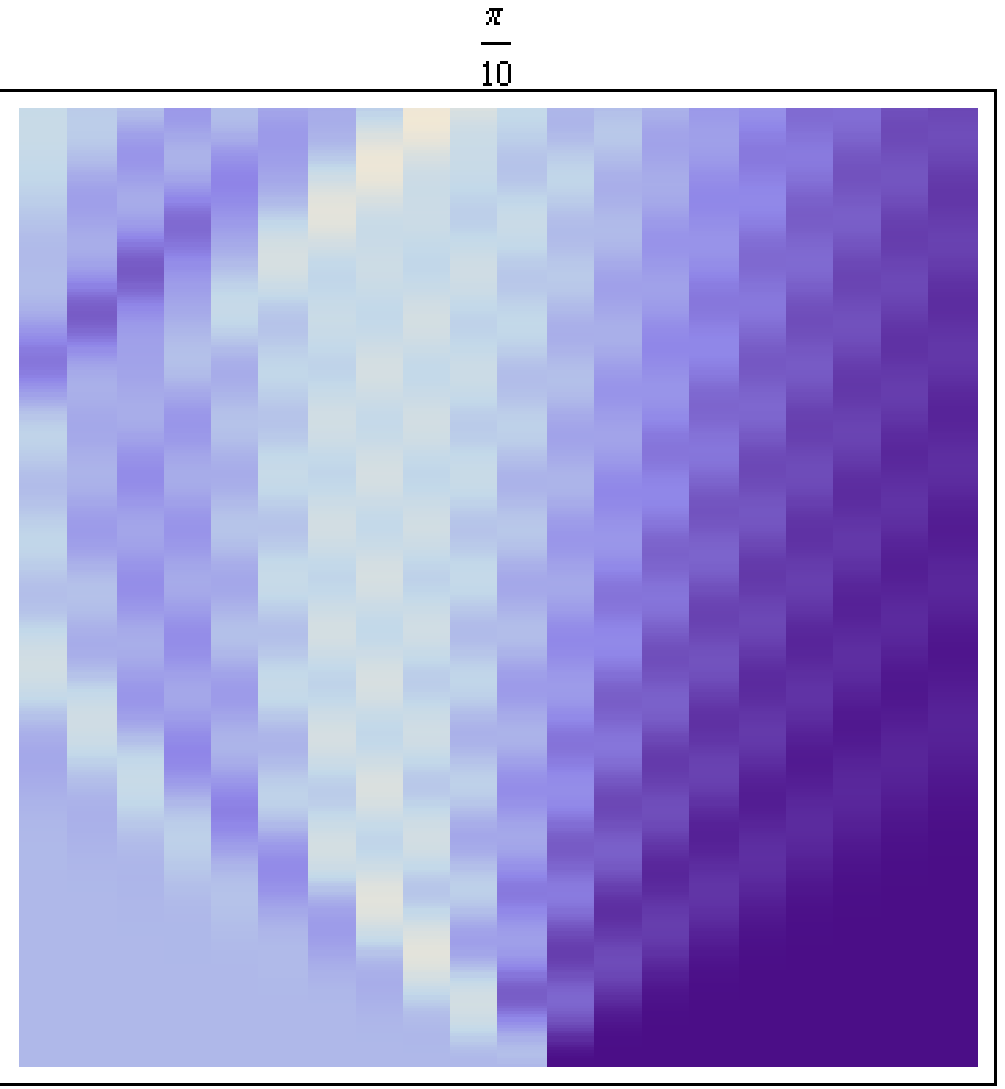} &  \includegraphics[scale=0.5]{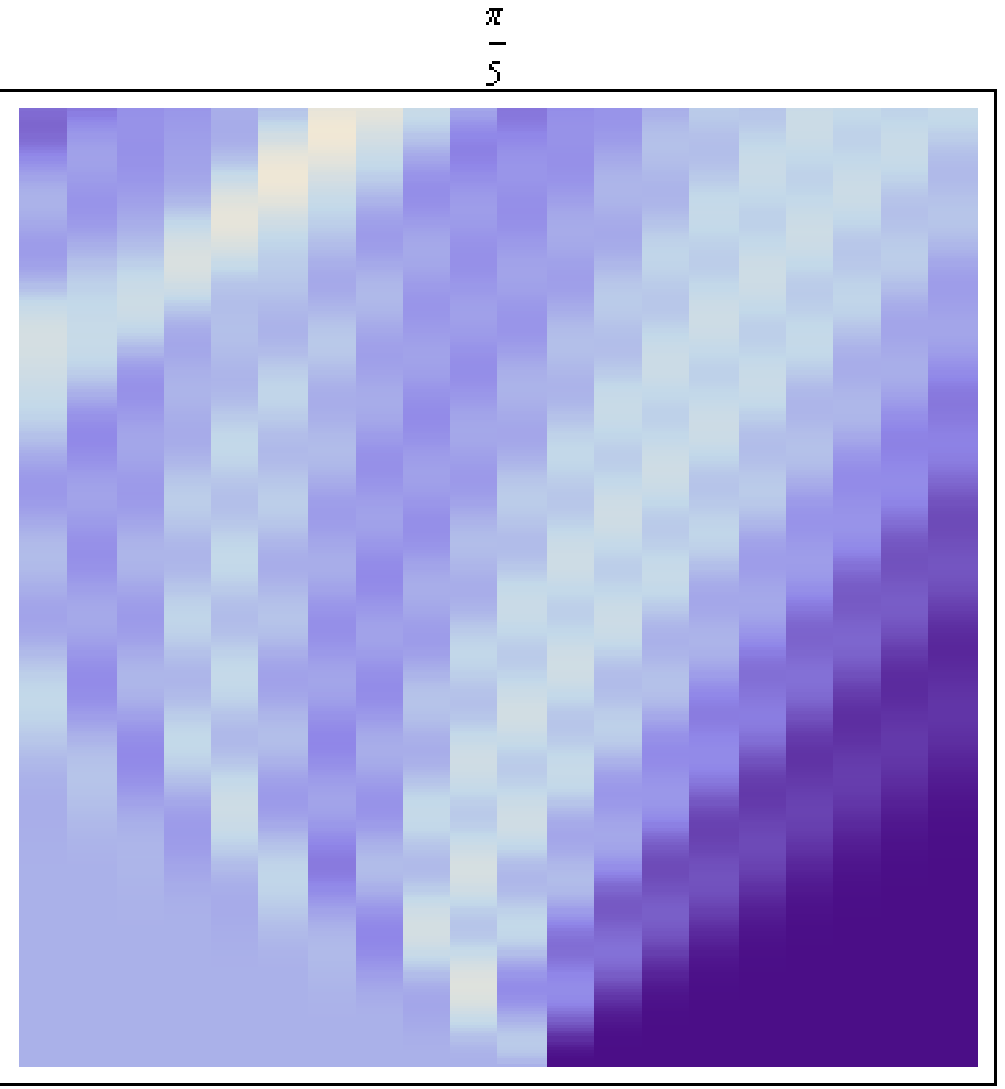} \\   \includegraphics[scale=0.5]{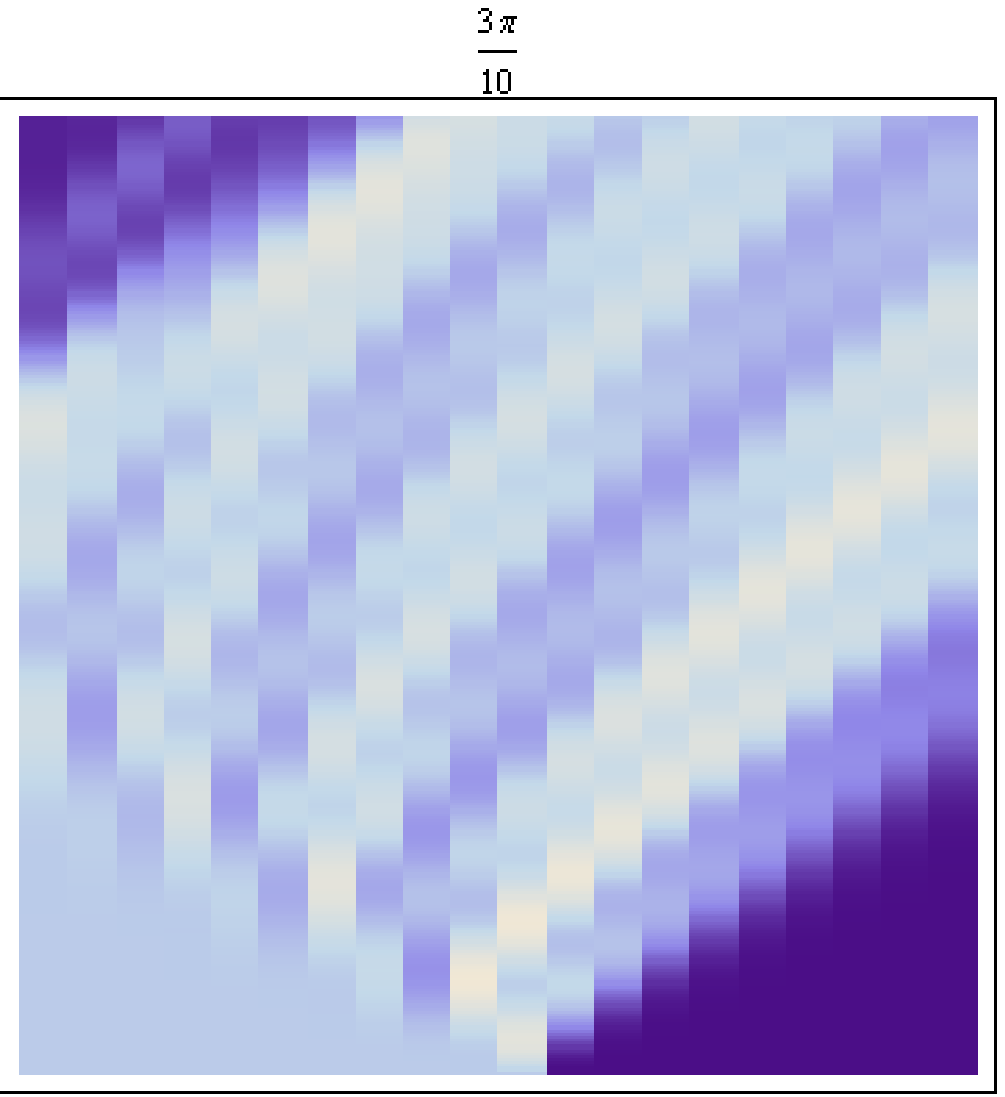} &  \includegraphics[scale=0.5]{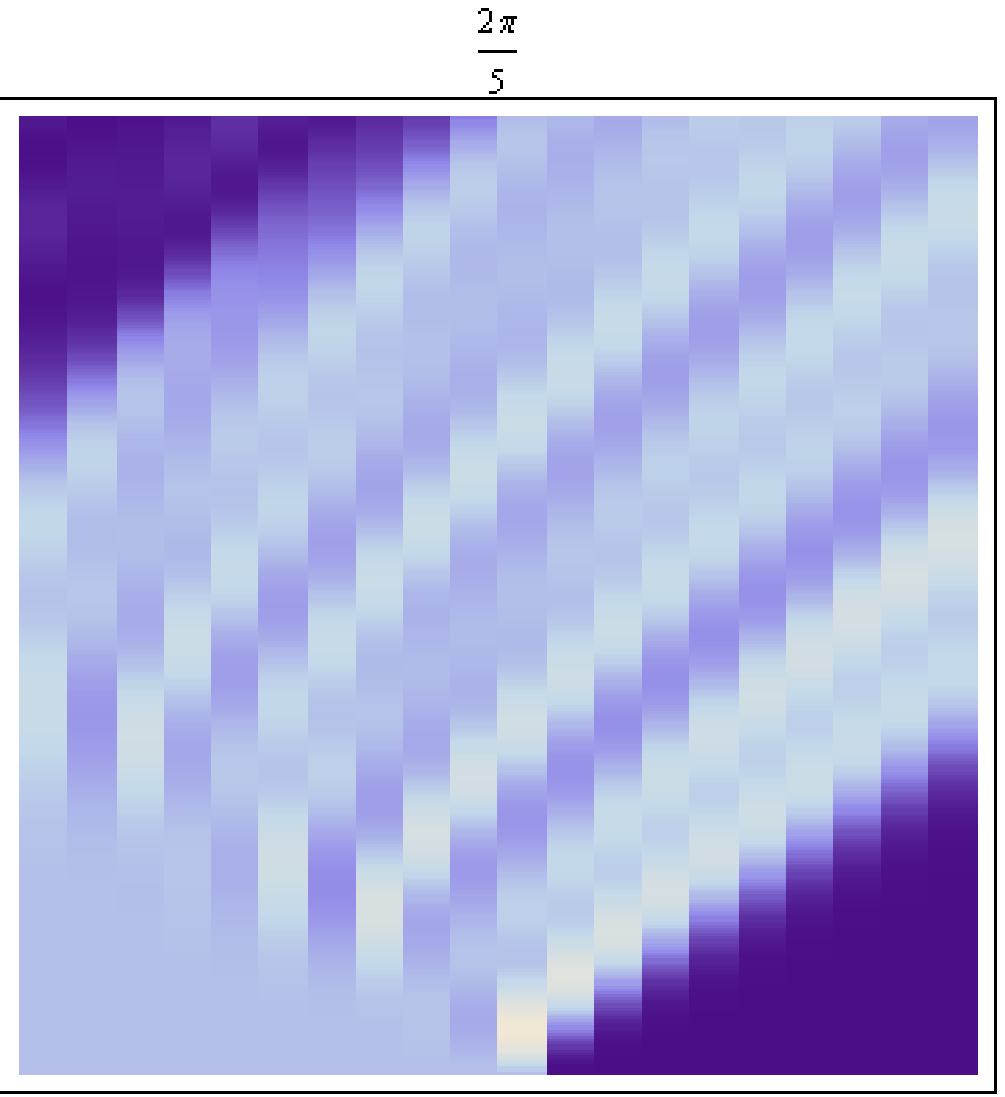} &  \includegraphics[scale=0.5]{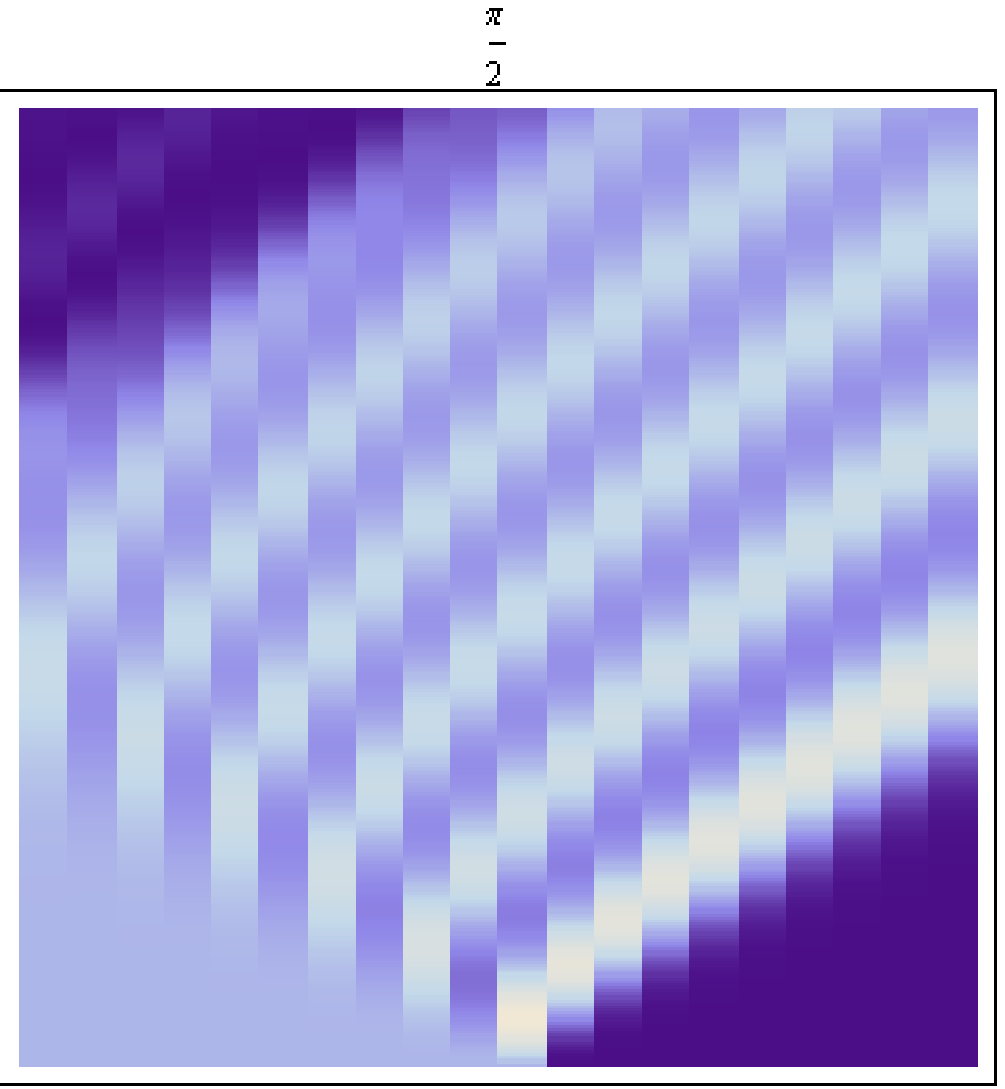}  \end{tabular} \end{center} 

\caption{ \label{fig:3} Probability density for the problem of diffraction in discrete space (abscissa) and continuous time (ordinate). In the first panel we show the position of the shutter denoted by S, and a pair of lines indicating the propagation of fronts at approximately constant velocity; they propagate both forward and backwards. The value of the Bloch quasi-momentum $k$ is varied for each panel and it is indicated on the top of each plot. The six shown patterns repeat in reversed order as $k$ increases beyond $\pi/2$. These effects can only happen in discrete space: An ever increasing $k$ does not lead to faster propagation. Notably, the structure outside the cone $t\pm 2n/e$ remains intact, as opposed to the richier structures found in the continuous variable solutions (Moshinsky functions). For this computation we have used the expansions in (\ref{e34}) up to 30 terms.} \end{figure}

A particular case of (\ref{e36}) ($q=\pi/2$) gives the possibility of extracting $\psi$ for $k=0$ in closed form:

\bea
\psi_{n}(t)|_{k=0} = i^{n} \left[\frac{   e^{it} + J_0 (t)}{2} - \sum_{m=0}^{n-1}   J_{m}(t) \right].
\label{e37}
\eea
Another limit of interest is the large time behaviour of $\psi$. This can be estimated in terms of the Meissel asymptotic form of Bessel functions as long as $(n,t)$ is not close to the caustic. We write $J_{n+m}(t) \approx \sqrt{\frac{2}{\pi t}} \exp \left( P_{n+m} + iQ_{n+m} \right)$ where $Q$ and $P$ are well-known functions (formulae (1) and (2) in p. 228 of \cite{1}). Upon substitution in (\ref{e34})

\bea
\psi_n(t) & \approx & i^{n}\sqrt{\frac{2}{\pi t}} \sum_{m=0}^{\infty} \exp \left( P_{n+m} + iQ_{n+m} \right).
\label{e38}
\eea
leading to the typical decay of the density as the inverse of $t$

\bea
|\psi_n(t)|^2 =\frac{2}{\pi t} |\sum_{m=0}^{\infty} \exp \left( P_{n+m} + iQ_{n+m} \right)|^2
\label{e39}
\eea
For a description of the probability density without approximations, see figure \ref{fig:3}.

\subsubsection{Maximal velocity in a non-relativistic framework}

This interesting property is a reminiscent of the behaviour seen in the point-like case. It can be derived fairly easily by considering short times and arbitrarily large index, such that each Bessel function in the expansion (\ref{e34}) can be replaced by the first term of the ascending series: 

\bea
\psi_n(t) = i^n \sum_{m=0}^{\infty} \left[ \frac{t^{n+m}}{(n+m)!} + \cdots\right]e^{-i(k-\pi/2)m}.
\label{e41}
\eea
Keeping lowest order in powers of $t$, the probability density can be approximated by

\bea
|\psi_n(t)|^2 \approx \frac{1}{(n!)^2} \left(\frac{t}{2}\right)^{2n},
\label{e42}
\eea
we can use the Stirling approximation to obtain constant probability density curves (say $|\psi|^2=\alpha$) in the form 

\bea
t = \pm \left(\frac{2 \alpha^{1/(2n)}}{e} \right) n.
\label{e43}
\eea
The cones described by these curves (straight lines for $\alpha=1$) are shown in the first panel of figure \ref{fig:3}. The presence of the cones is independent of the quasi-momentum $k$. This property is found exclusively in discrete problems, as non-relativistic quantum mechanical systems in continuous variables do not suffer the restrictions of a maximal velocity of propagation. 

\subsubsection{Circular property of the solution}

\begin{figure}[!h] \begin{center} \begin{tabular}{ccc} \includegraphics[scale=0.5]{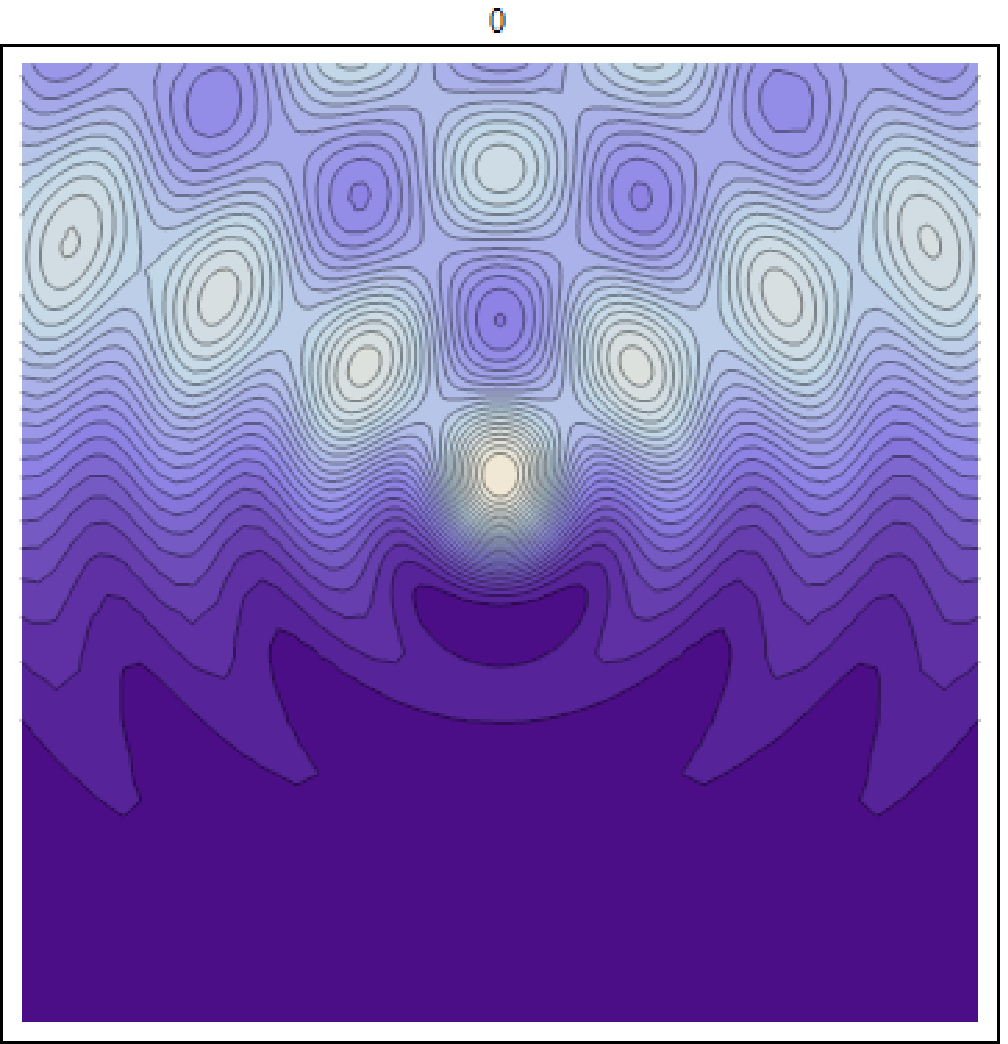} &  \includegraphics[scale=0.5]{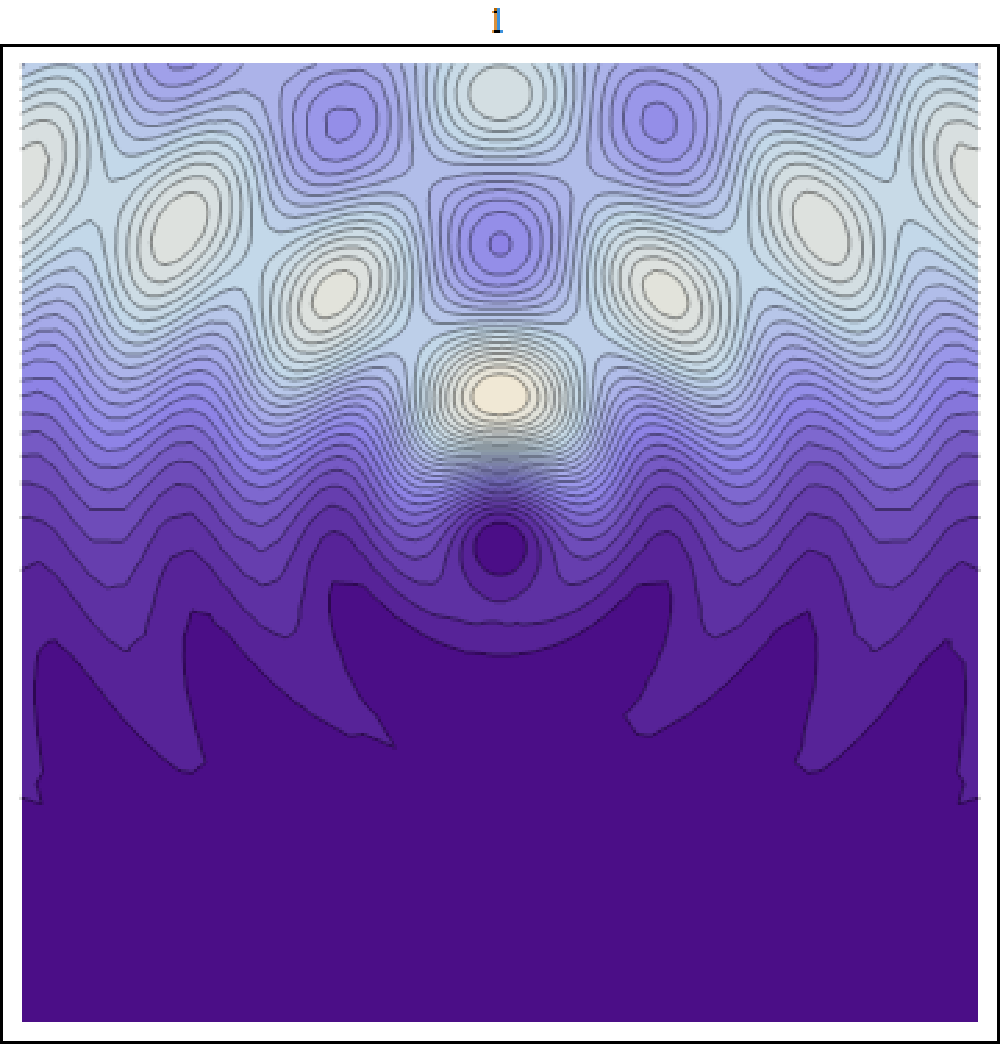} &  \includegraphics[scale=0.5]{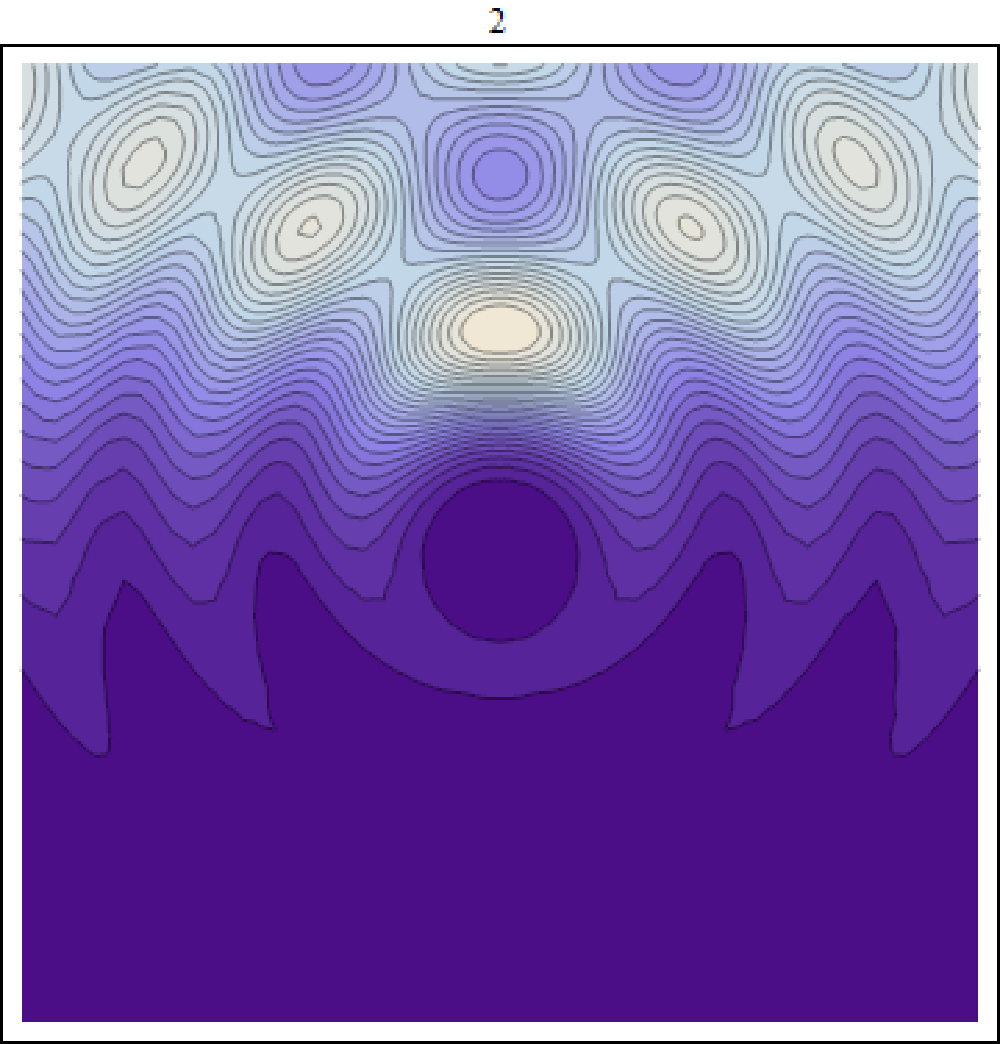} \\   \includegraphics[scale=0.5]{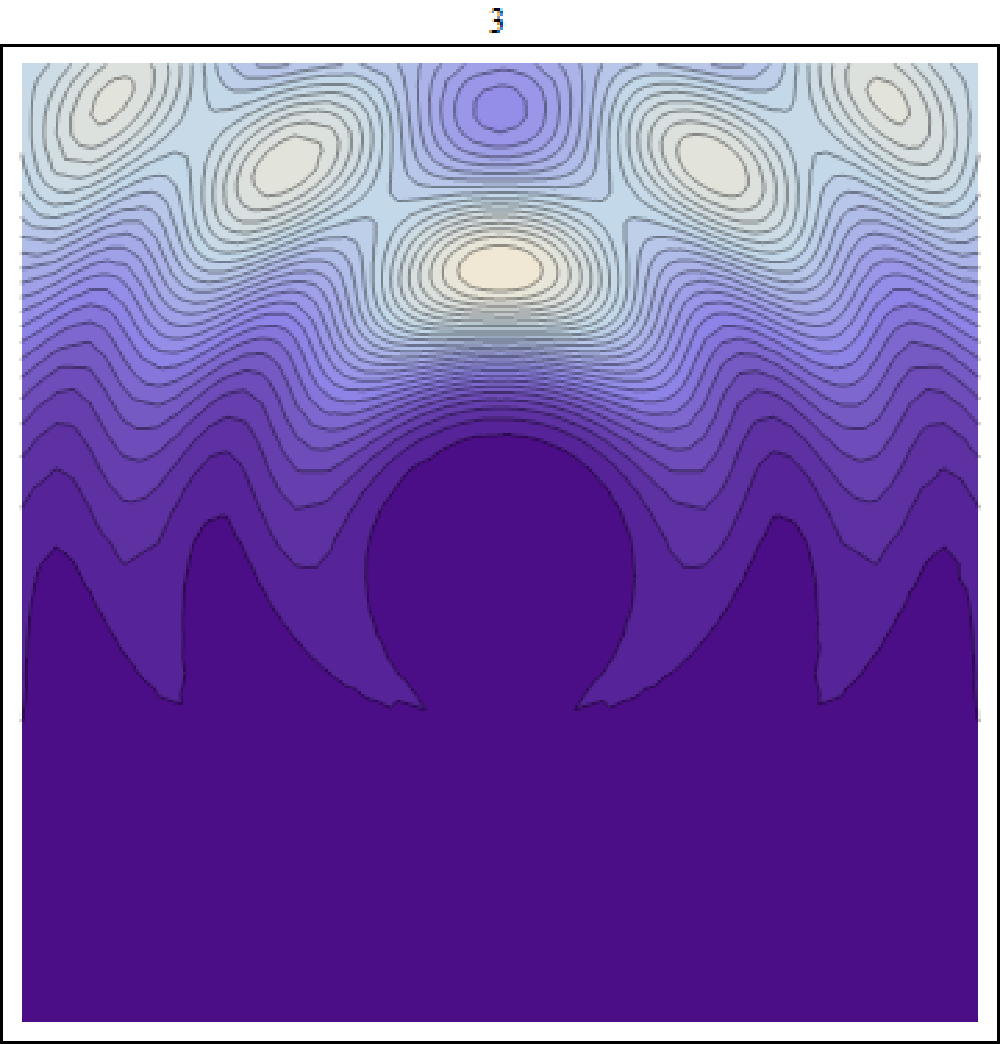} &  \includegraphics[scale=0.5]{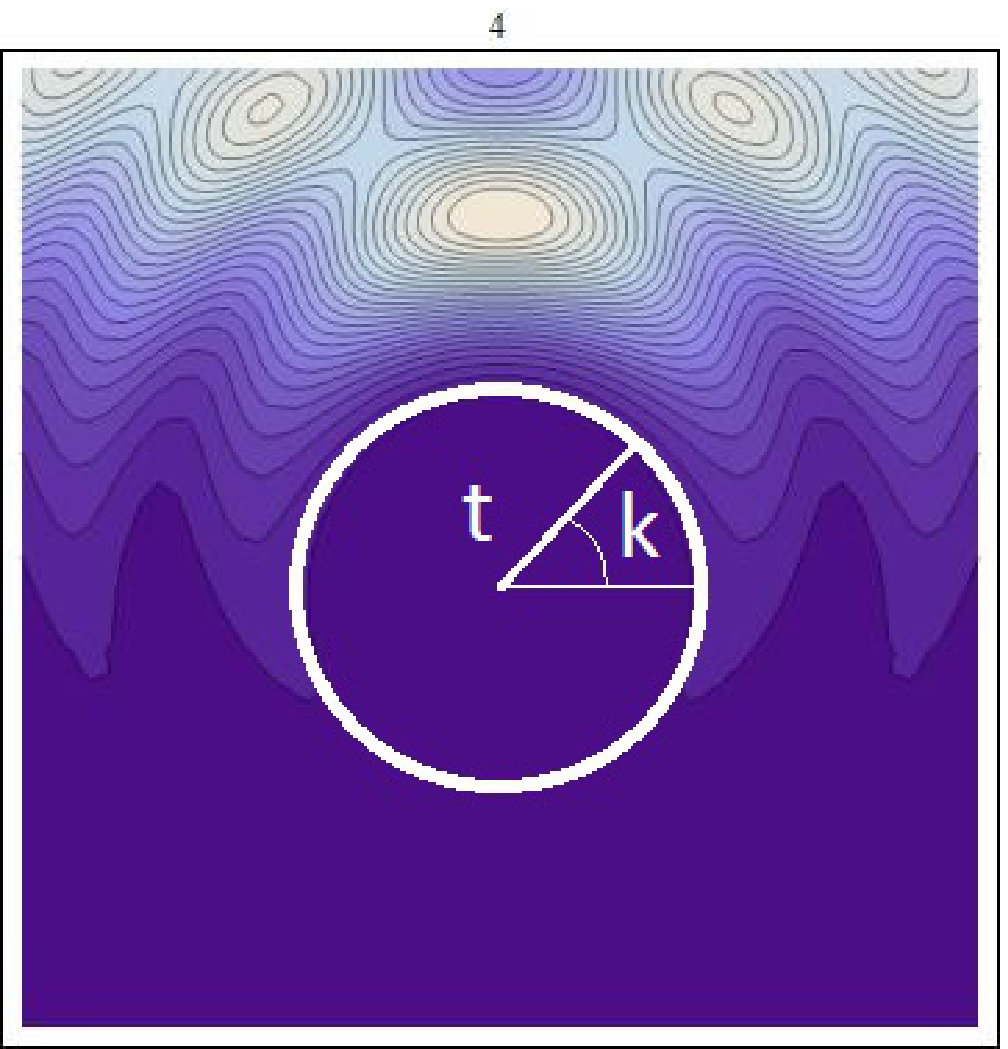} &  \includegraphics[scale=0.5]{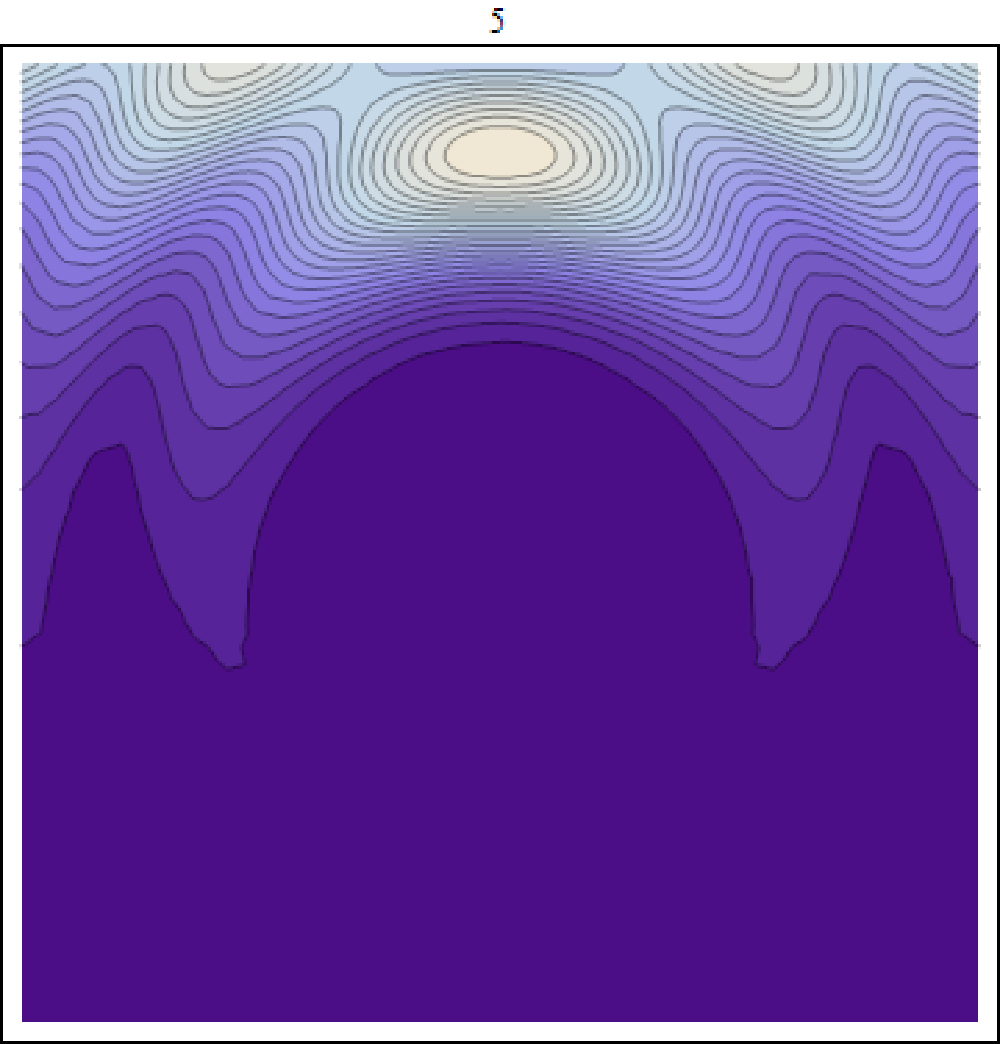}  \end{tabular} \end{center} 

\caption{ \label{fig:4} Diffraction in discrete space as a circular wave. Each panel represents the probability at a fixed site, going from dark areas (low probability) to bright areas (high probability). The radius stands for $t=\sqrt{x^2+y^2}$, while the angle denotes the quasi-momentum $k=\arctan (y/x)$. On top of each plot we indicate the value of the position $n$. A circular pattern around the origin emerges, showing that a signal arrives at the specified point after a fixed time, irrespectively of the initil momentum of the wave. One of such circles is depicted in the central panel of the lower row. } \end{figure}

The time-diffractive solution (\ref{e35}) can be related to circular waves. This property can be recognized by inspecting the expansion (\ref{e34}). When we replace the time and quasi-momentum variables by radius and angle in the form $t \mapsto r, k \mapsto \varphi$, we immediately see that

\bea
\psi_n(r) = e^{in\varphi} \sum_{m=0}^{\infty} \Phi_{m+n}(r,\varphi) 
\label{e51}
\eea
where $\Phi_{l}(r,\varphi)$ is a circular wave of angular momentum $l \in \textbf{Z}$ and unit energy $E=1$. This means that the function $\Psi_{n}(r,\varphi) \equiv e^{-in\varphi} \psi_n(r) $ satisfies the Helmholtz equation of unit wave vector in polar coordinates $x = t \cos k, y = t \sin k$, namely

\bea
\nabla_{x,y}^2 \Psi_{n}(r,\varphi)  + \Psi_{n}(r,\varphi) = 0, \qquad \forall n
\label{e52}
\eea
and it corresponds to a free wave in two-dimensional space with a non-vanishing boundary condition at the origin:

\bea
\Psi_{l}(0,\varphi) = \cases{1 \quad \textrm{for} \quad l<0 \\ 0 \qquad \textrm{otherwise}}
\label{e53}
\eea
\ie  a left-handed solution, since positive angular momenta do not contribute in (\ref{e53}). A depiction of the diffractive process in terms of circular wavefronts is shown in figure \ref{fig:4}.

\subsubsection{Recursion in the position variable and algebraic properties}

Since our time-diffractive solution is related to circular waves, we expect a number of well-known properties to be applicable in our study, in particular when we try to evaluate the wave function at a certain position by knowing its value at a distant point. We can extract a recursion relation in $n$ by noting that a circular wave of arbitrary angular momentum can be obtained by ``cranking up" the solutions of lower angular momenta. In the language of Bessel functions, this corresponds to a ladder operator which is linear in the radial derivative. For the full circular wave, this process can be generated by a chiral operator. We have

\bea
\Phi_{n\pm1}(r,\varphi) &=& -e^{\pm i\varphi} \left[\frac{i}{r} \partial_{\varphi} \pm \partial_{r} \right] \Phi_{n}(r,\varphi)
\label{e54}
\eea
leading to the validity of (\ref{e54}) for any superposition of $\Phi$'s, in particular

\bea
\Psi_{n+1}(r,\varphi) &=& -e^{ i\varphi} \left[\frac{i}{r} \partial_{\varphi} + \partial_{r} \right] \Psi_{n}(r,\varphi)
\label{e55}
\eea
Its application to $\Psi_0$ an arbitrary number of times gives

\bea
\Psi_{n}(r,\varphi) &=& \left\{ -e^{ i\varphi} \left[\frac{i}{r} \partial_{\varphi} + \partial_{r} \right] \right\}^n \Psi_{0}(r,\varphi).
\label{e56}
\eea
Finally, using the definition of $\Psi$ in terms of $\psi$, we obtain a new representation of the wavefunction at distant points in terms of the solution at the shutter:

\bea
\psi_{n}(t)|_k &=& e^{ikn} \left\{ -e^{ ik} \left[\frac{i}{t} \partial_{k} + \partial_{t} \right] \right\}^n \psi_{0}(t)|_k \quad.
\label{e56}
\eea
With this, we establish that ``hopping'' over discrete sites provides means to evaluate diffractive solutions. This process is related to the algebraic properties of the problem, obtained from the factorization of $\nabla_{t,k}^2$.

\subsection{An edge producing a Bessel profile}

Many other examples of evolution can be studied in exact form by using several results on the closed summation of Bessel functions. In connection with diffraction problems by edges, we envisage a wide class of results by means of degenerate addition theorems where the summation index starts at $0$. Lommel expansions constitute another possibility. Take, for instance, an initial condition in the form of a {\it ramp\ }:

\bea
\psi_n (0) = \theta(n) \frac{b^n}{n!}.
\label{e61}
\eea
 This {\it poissonian\ }profile represents the wave function of a particle that is trapped near the shutter, with a fastly decreasing tail. The function vanishes abruptly for negative positions and it can be regarded as a member of the family of edges. The real parameter $b$ determines the position of the maximum, e.g. for $b=1$ the maximum is exactly at the position of the edge. The evolution is obtained by computing the sum

\bea
\psi_n(t) &=& \sum_{m=0}^{\infty} \frac{b^m}{m!} i^{n-m} J_{n-m}(t) \nonumber \\
&=&\left[ i^{n} (t-2ib)^{n/2} \right] \times \left[ t^{-n/2}J_n(\sqrt{t^2-2ibt}) \right]
\label{e62}
\eea
where the second factor is manifestly regular at $t=0$. In this derivation we have used the Lommel expansion (formula (5) p. 141 \cite{1})

\bea
J_{\nu}(z\sqrt{1+w}) = (1+w)^{\nu/2}\sum_{m=0}^{\infty}\frac{(-\frac{wz}{2})^m}{m!}J_{\nu+m}(z) .
\label{e63}
\eea
From this result we can infer many properties that coincide with our previously obtained effects. For instance, using the first term in the ascending series of $J$ one can find constant probability curves involving $n$ and $t$, finding a short time behaviour of the caustics and a maximal velocity effect. A decreasing density in the form $1/t$ can be found from the asymptotic form of $J^2$. When $b=0$, we recover the point-like source. 

\section{Scattering by an edge in discrete space as a problem of propagation}

It is well-known that the shutter problem in continuous variables represents a diffraction problem in two spatial dimensions $-$Fresnel diffraction. Is it possible to show a similar analogy between our discrete problem and a scattering system in two dimensions? In the rest of this paper we clarify the connection of our evolving edges with a genuine two-dimensional problem described by the Helmholtz equation. To this end, we demonstrate how a problem of scattering with a periodic structure as a background (one discrete variable and one continuous variable) can be written in terms of a propagator. The corresponding kernels can be given in explicit form.

\subsection{Propagator for edge diffraction in two continuous variables}

For the sake of completeness, let us deal first with the two-dimensional {\it continuous\ }problem of a scalar plane wave scattered by an edge. The problem can be written in terms of forward and backward propagation (transmission and reflection) in the following manner: Suppose that a plane wave arrives at a screen positioned at $z=0$ and parallel to the $x$ axis (at this point, we make no further assumptions on the nature of such a physical obstacle). We have $\psi(z<0,x) = \exp (i k_z z + i k_x x)$ with $k_z, k_x$ the corresponding components of the momentum. At $z=0$ we can impose an arbitrary boundary condition determined by the form of the screen; for definiteness we may put $\psi(0,x)= \theta(-x)e^{ixk_x}$. Set the energy $E=k_x^2 + k_z^2$ such that

\bea
\left[\nabla_{z,x}^2 + E \right] \psi(z,x)=0.
\label{e71}
\eea
Since the solution of the problem satisfies the Helmholtz equation, it can be written formally as a continuous superposition of the form

\bea
\fl \psi(z,x) = A_{\rm{R}} \int_{-\infty}^{\infty} dk \phi_k^{\rm{R}} \exp \left(ikx+iz\sqrt{E-k^2} \right) + A_{\rm{L}} \int_{-\infty}^{\infty} dk \phi_k^{\rm{L}} \exp \left(ikx-iz\sqrt{E-k^2} \right) \nonumber \\ \quad
\label{e72}
\eea
where L and R indicate solutions propagating to the left and to the right, respectively. Now, let us assume no reflection ($ A_{\rm{L}}=0,  A_{\rm{R}}=1 $) 

\bea
 \psi(z,x) =  \int_{-\infty}^{\infty} dk \phi_k \exp \left(ikx+iz\sqrt{E-k^2} \right)
\label{e73}
\eea
and let us evaluate $\psi$ at $z=0$ in order to determine the distribution in $k$:

\bea
\psi(0,x)=  \int_{-\infty}^{\infty} dk \phi_k e^{ikx} \Rightarrow \phi_k= \frac{1}{2\pi} \int_{-\infty}^{\infty} dk \psi(0,x) e^{-ikx}.
\label{e74}
\eea
Substituting the inverse Fourier transform of the initial condition we have

\bea
 \fl \psi(z,x) = \frac{1}{2\pi}\int_{-\infty}^{\infty} dx' \psi(0,x') \left[ \int_{-\infty}^{\infty} dk \exp \left(ik(x-x')+iz\sqrt{E-k^2} \right) \right].
\label{e75}
\eea
Finally, the propagator in fictitious time $z$ and parameter $E$ is given by the factor in brackets

\bea
 K_E(x,x';z) = \frac{1}{2\pi} \int_{-\infty}^{\infty} dk \exp \left(ik(x-x')+iz\sqrt{E-k^2} \right).
\label{e76}
\eea
This integral can be evaluated upon a change of variables \cite{sym} and the appropriate choice of an integration contour. Let $\eta=E^{-1/2}(k+i\sqrt{E-k^2})$ and $\eta^{-1}=E^{-1/2}(k-i\sqrt{E-k^2})$, with an integration line given by the segments $\left\{\rm{Re}(\eta) < -1 \right\} \cup \left\{ \eta=e^{i\delta} | 0<\delta< \pi \right\} \cup \left\{ 1>\rm{Re}(\eta ) >0\right\} $. As there are no poles in the way, the upper semi-circular curve and the segment $(0,1)$ can be continuously deformed such that $\eta \in (-\infty,0)$. The integral in (\ref{e76}) is transformed to 

\bea
 \fl K_E(x,x';z) = \frac{\sqrt{E}}{4\pi} \int_{-\infty}^{0}d\eta \left[1-\eta^{-2}\right] \exp\left\{ \frac{i\sqrt{E}}{2} \left[ \eta(x-x'+iz)+\eta^{-1}(x-x'-iz) \right] \right\} \nonumber \\
\fl =  \frac{\sqrt{E}}{4\pi}  \int_{0}^{\infty} d\eta \left\{ \eta^{-2}  e^{ \frac{i\sqrt{E}}{2} \left[ \eta(x-x'+iz)+\eta^{-1}(x-x'-iz) \right] } -   e^{ \frac{i\sqrt{E}}{2} \left[ \eta(x-x'+iz)+\eta^{-1}(x-x'-iz) \right] } \right\}.
\label{e77}
\eea
The closed result is
\bea
K_E(x,x';z) =\left[\frac{-iz\sqrt{E}}{2\sqrt{(x-x')^2+z^2}} \right]  H_1^{(1)}\left( \sqrt{E} \sqrt{(x-x')^2+z^2} \right),
\label{e78}
\eea
where we have used the integral representation of the Hankel functions $H_{\pm1}^{(1)}$ \cite{3} and trivial algebraic manipulations. The result found here should not be confused with the usual Green's function of the 2D Helmholtz equation subject to Sommerfeld's outgoing radiation condition (given by the Hankel function $H_0^{(1)}$ alone). Our boundary condition has been defined along the $x$ axis at $z=0$.

 It is important to note that in (\ref{e78}), the two limits $\sqrt{(x-x')^2+z^2}\gg 1/\sqrt{E}$ (short wavelengths) and $(x-x')^2/z^2 \ll 1$ (paraxiality) must be taken {\it simultaneously\ }in order to recover the free Schr\"odinger propagator by means of the asymptotic form of the Hankel function:

\bea
 K_E(x,x';z) \approx \left[e^{iz\sqrt{E}}\right] \times \left(\frac{\sqrt{E}}{2\pi i z}\right)^{1/2} \exp\left( \frac{i \sqrt{E}(x-x')^2}{2z} \right).
\label{e79}
\eea
The prefactor in brackets is related to the planewave propagating originally to the right on the left semi-plane $z<0$, with a wavelength $\lambda= 2\pi/\sqrt{E}$. 

Although the problem of diffraction by an edge has been discussed exhaustively in the literature, no description of the propagator (\ref{e78}) can be found in previous works. We should stress that this is the Euclidean version of a relativistic problem described by the Klein-Gordon equation \cite{sym}. Since the solutions to relativistic diffraction in time were identified as Lommel functions \cite{2}, we should expect similar results in our Euclidean case. 

In connection with the validity of the approximations leading to the Schr\"odinger kernel, it is worth to mention that the problem of diffraction in propagation axis $z$ is characterized by infinitesimal oscillations near $z=0$ due to a dependence of the form $(x-x')^2/z$ in (\ref{e79}). On the other hand, the kernel (\ref{e78}) exhibits a dependence on $\sqrt{(x-x')+z^2}$ which is regular at $z=0$ and does not produce infinitesimal oscillations. Now that we have set the continuous example, we move to the discrete problem.

\subsection{Continuous propagation axis and discrete transverse dimension}

\begin{figure}[!h] \begin{center} \begin{tabular}{cc} \includegraphics[scale=0.5]{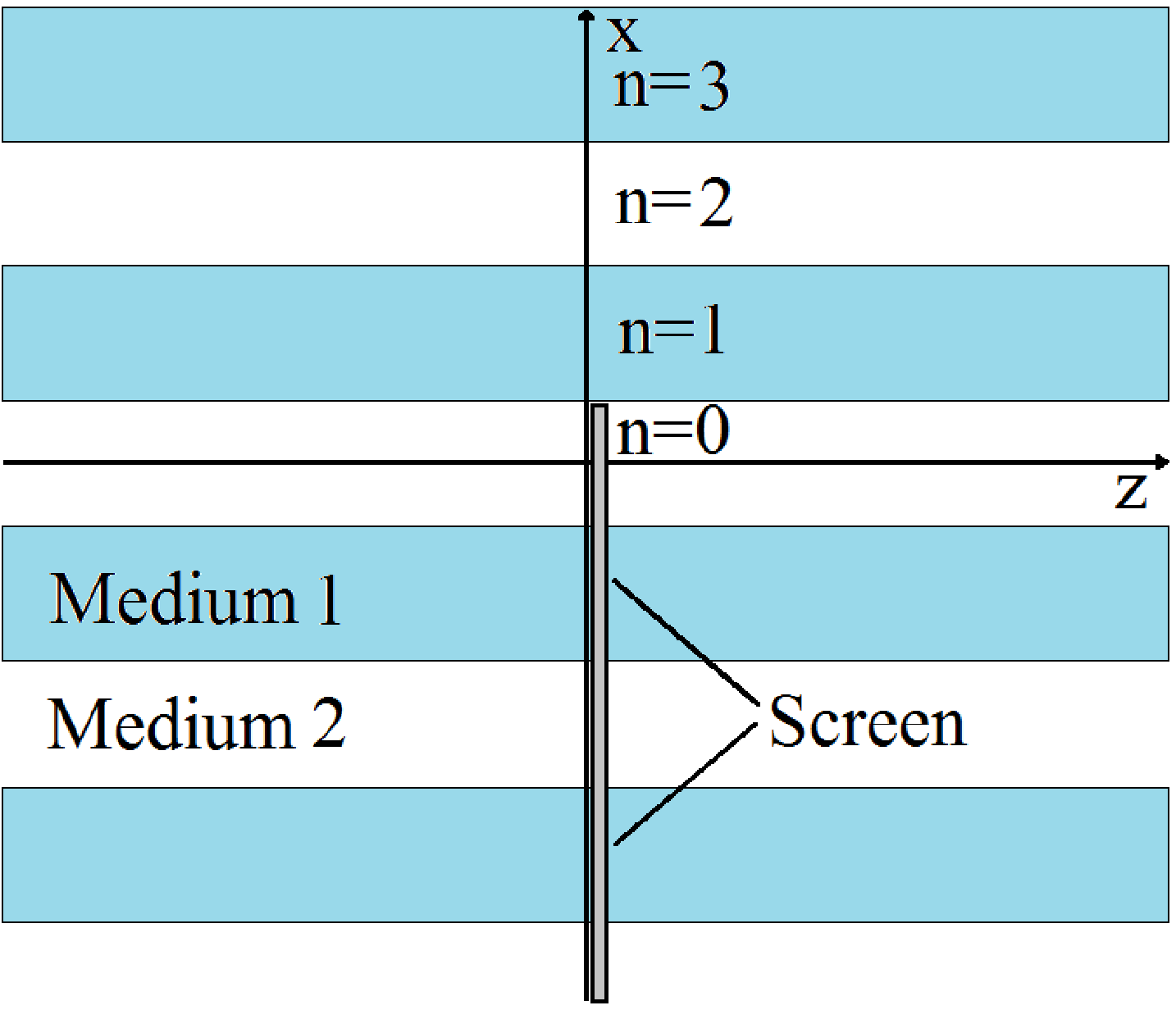} &  \includegraphics[scale=0.5]{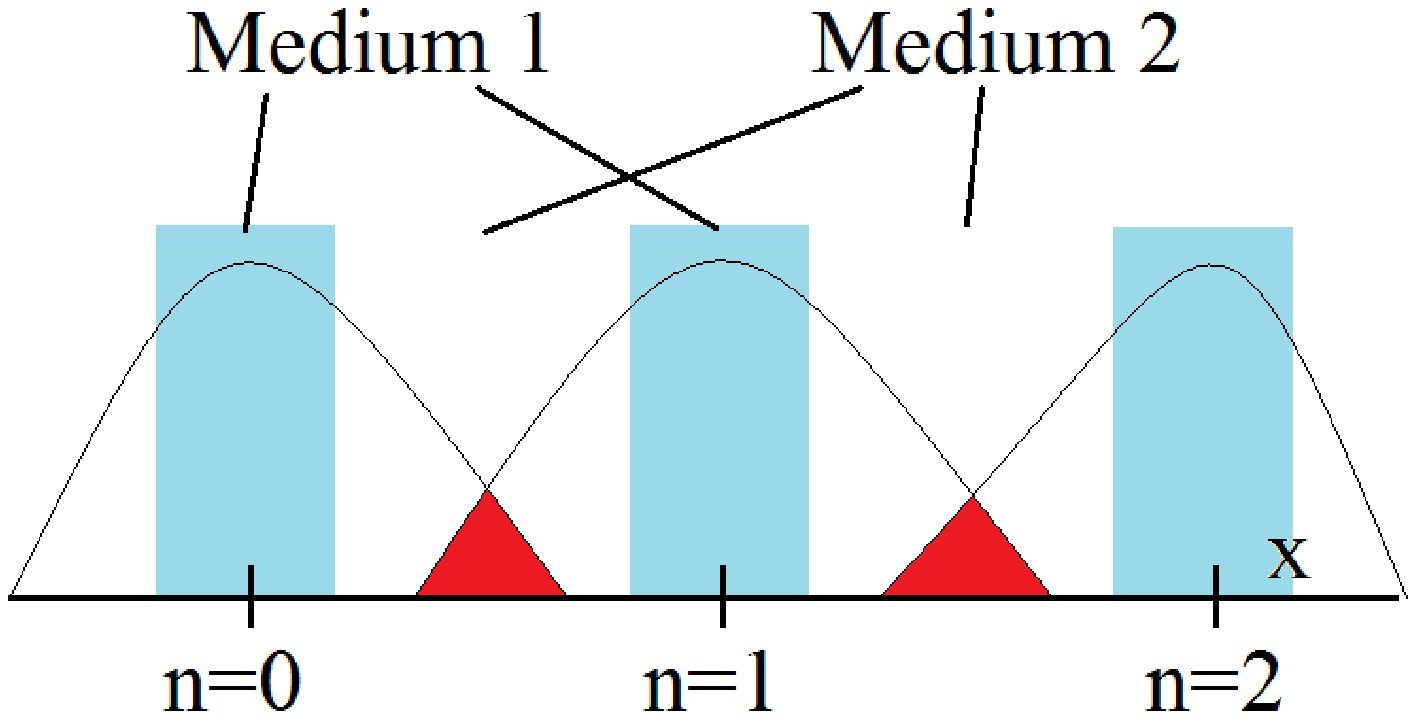} \end{tabular} \end{center} 

\caption{ \label{fig:5} Left panel: A periodic background realized through the alternation of two materials (solid state) or two potentials (quantum case). A screen blocks a wave propagating along $z$. Right panel: The two media represented as potential barriers along the $x$ variable. A sketch of the profile for localized states is shown, with coloured areas indicating the overlaps and nearest-neighbour interactions.} \end{figure}

We turn our attention to the Helmholtz equation of one continuous variable $z$ and one discrete variable $n$. This problem also corresponds to a wave propagating in a 2D medium with one periodic direction (see section \ref{sec:5} and figure \ref{fig:5}). The discrete derivative operator along the variable $n$ contains twice the identity operator, therefore it can be simplified by a redefinition of the energy in the full Helmholtz equation, leading to the analog of the l.h.s. of (\ref{e2}). We have

\bea
\frac{\partial^2 \psi_n(z)}{\partial z^2} + \psi_{n+1}(z) + \psi_{n-1}(z) + E \psi_n(z) = 0.
\label{e81}
\eea

In complete analogy with the continuous 2D example, we state our boundary problem as follows:  A Bloch wave propagates in the left semi-plane towards an absorptive edge, \ie $\psi_n(z<0)=e^{ik_x n + ik_z z}$, where $k_x$ is Bloch's quasi-momentum and $k_z$ is the momentum corresponding to coordinate $z$. The energy $E= k_z^2 - 2\cos k_x$ is fixed. The most general superposition which solves (\ref{e81}) when the wall is absorptive can be written as

\bea
\psi_n(z) = \int_{0}^{2\pi} dk \phi_k \exp \left( ik n + i z \sqrt{E+2\cos k} \right), \qquad z>0
\label{e82}
\eea
where the integration over $k$ extends to the first Brillouin zone $(0,2\pi)$. The boundary condition at the position of the screen implies

\bea
\psi_n(0) =  \int_{0}^{2\pi} dk \phi_k e^{ ik n} \Rightarrow \phi_k = \frac{1}{2\pi}  \sum_{m \in \textbf{Z}} e^{- ik m}\psi_m(0).
 \label{e83}
\eea
For definiteness, we may choose 

\bea
\fl \psi_n(0) = e^{ik_x n} \theta(-n), \qquad \phi_k =  \frac{1}{2\pi} \sum_{m=0}^{\infty} e^{i(k-k_x)m} = \frac{e^{i(k_x-k)/2}}{2\pi i \sin \left[ (k_x-k)/2 \right]}
\label{e84}
\eea
although the propagator we are seeking for does not require such an explicit function. Using the Fourier series in (\ref{e83}) for $\phi_k$, we have

\bea
\psi_n(z) = \frac{1}{2\pi} \sum_{m \in \textbf{Z}}\psi_m(0) \left[ \int_{0}^{2\pi}dk e^{ik(n-m)+iz\sqrt{E+2\cos k}} \right]
\label{e85}
\eea
and, as before, the propagator can be read-off from the expression between brackets:

\bea
K_E(n,m;z) =  \frac{1}{2\pi}  \left[ \int_{0}^{2\pi}dk e^{ik(n-m)+iz\sqrt{E+2\cos k}} \right].
\label{e86}
\eea
The discrete kernel (\ref{e86}) can be expressed in compact form by employing a trick used by Dattoli \cite{1.2} (with reminiscence from a formula by Filon, see formula (5) p. 51 \cite{1}). A translation operator of variable $E$ and parameter $2\cos k$, acting on the function $e^{i z \sqrt{E}}$  can be expanded in terms of Bessel functions as follows:

\bea
\fl e^{i z \sqrt{E+2\cos k}}=\exp \left[ i\cos k \left(-2i\frac{\partial}{\partial E}\right)\right] e^{i z \sqrt{E}} = \left[ \sum_{q \in \textbf{Z}} i^q e^{iqk} J_q \left(-2i\frac{\partial}{\partial E} \right) \right] e^{iz\sqrt{E}}
\label{e87}
\eea
where each Bessel operator is understood in terms of its ascending series. Inserting back in (\ref{e86}) and performing the integration over $k$ one finally gets the simple expression

\bea
K_E(n,m;z)= i^{n-m} J_{n-m}\left(-2i\frac{\partial}{\partial E} \right) e^{iz\sqrt{E}}
\label{e88}
\eea
which ressembles our discrete Schr\"odinger propagator (\ref{e7}) in the form

\bea
K_E(n,m;z)=K_{\textrm{\scriptsize Schr\"odinger}} \left(n, m; -2i\frac{\partial}{\partial E} \right) e^{iz\sqrt{E}}.
\label{e89}
\eea
This operational form has several uses, according to our needs \cite{note2}. For example, consider the following identity for some analytic $f$

\bea
f\left(-2i\frac{\partial}{\partial E} \right) e^{iz\sqrt{E}} =  e^{iz\sqrt{E}} \left[f\left(-2i\frac{\partial}{\partial E} + \frac{z}{\sqrt{E}} \right) \right] \left[ 1 \right]
\label{e90}
\eea
where $1$ is the constant function equal to unity. In Appendix B we show that this identity can be approximated (in a short wavelength regime) by the expression

\bea
f\left(-2i\frac{\partial}{\partial E} \right) e^{iz\sqrt{E}} \approx  e^{iz\sqrt{E}} \left[f\left(\frac{z}{\sqrt{E}} \right) + \frac{iz}{4E^{3/2}} f''\left(\frac{z}{\sqrt{E}} \right) \right].
\label{e91} 
\eea
 This allows to compute the first corrections to our propagator $K_E$ by replacing the discrete Schr\"odinger kernel $K$ in $f$ and $f''$:

\bea
\fl K_E(n,m;z) \approx   e^{iz\sqrt{E}} \left[ K \left( n,m;\frac{z}{\sqrt{E}} \right) \right. \nonumber \\ \fl - \left. \frac{iz}{16E^{3/2}} \left\{ K\left( n+2,m;\frac{z}{\sqrt{E}} \right)+ K \left( n-2,m;\frac{z}{\sqrt{E}} \right)+2K \left( n,m;\frac{z}{\sqrt{E}} \right) \right\} \right]. 
\label{e92}
\eea
The first term in (\ref{e92}) establishes the correct limit for short wavelengths. It gives the discrete Schr\"odinger theory with fictitious time $t = z/\sqrt{E}$ and the additional plane wave factor  $e^{iz\sqrt{E}}$. For diffraction problems, we can establish a similar approximation for the wave function in terms of the Schr\"odinger solution $\psi$ in (\ref{e35})

\bea
\fl \psi_n(z,E) \approx  e^{iz\sqrt{E}} \left[ \psi_n \left(\frac{z}{\sqrt{E}} \right) - \frac{iz}{16E^{3/2}} \times \right. \nonumber \\ \left. \times \left\{ \psi_{n+2}\left( \frac{z}{\sqrt{E}} \right)+ \psi_{n-2} \left( \frac{z}{\sqrt{E}} \right)+2 \psi_n \left( \frac{z}{\sqrt{E}} \right) \right\} \right]
\label{e93}
\eea
This is our final result: The wave function of a diffractive experiment in 2D ressembles, in a suitable approximation, the problem of diffraction in discrete space and continuous time. The parallelism {\it Shutter\ }$\leftrightarrow${\it Fresnel \ }between the continuous problems is also valid for our discrete settings. Moreover, the effect is accompanied by the short wavelength corrections analogous to those separating the Hankel function (\ref{e78}) from the gaussian kernel (\ref{e79}). This has consequences in the propagation of waves on a periodic background (for example, a photonic structure).

\section{Connection with tight-binding arrays and photonic structures \label{sec:5}}

The connection between the discrete version of the Schr\"odinger equation and tigh-binding arrays is rather direct. Let us focus in the lowest energy band of a homogeneous polymer chain with nearest neighbour interactions. In the appropriate units, the hamiltonian for this band is 

\bea
H= -\frac{1}{2} \sum_{n \in \textbf{Z}}\left[|n\> \<n+1| + |n+1\> \<n|\right]
\label{e101}
\eea
where any localized wavefunction $\xi$ at site $n$ can be obtained via the kets above: $\xi_n(x)=\<x|n\>$. Our present work on the propagator leads to the solution of the time dependent-problem described by the wave function
\bea
\psi(x,t) = \sum_{n \in \textbf{Z}} \psi_n(t) \xi_n(x),
\label{e102}
\eea
with the continuous-variable propagator restricted to one band given by

\bea
K_{\textrm{\scriptsize polymer}}(x,x';t) = \sum_{n,m \in \textbf{Z}} \xi_n(x) \xi^*_m(x') K(n,m;t).
\label{e103}
\eea
In a polymeric problem, we are usually given the localized functions $\xi_n(x)$ describing a quantum particle at isolated sites. When we deal with the propagation of some initially known packet $\psi_0(x)$, we assume that the packet can be completely resolved by the basis of localized states $\psi_0(x) = \sum_n C_n \xi_n(x)$ (Wannier). In other words, we rely completely on the information given by the overlap $\int dx \psi_0(x) \xi^*_n(x)$. This is the first step towards a time-dependent description of conductive (conjugate) polymers of a more complicated nature \cite{9}.

%In a similar manner, but in physics at a different scale, we can describe the scattering of an electromagnetic wave from a screen with a periodic structure as a background. This can be %realized by working with TM modes in a quasi 2D cavity \cite{8} with dielectric inclusions in the form of stripes. The scalar wave is given by one component of the electric field. The frequency %of the calculated band should lie below the fundamental frequency determined by the distance between top and bottom plates of the quasi 2D cavity.  

\section{Summary}

In this paper we have calculated a propagator in discrete variables, with no precedent in the standard literature \cite{3.1, 3.2, 3.3}. In order to understand such a novel object, we have studied its properties and extended the Feynman path sums to discrete variables. We discussed some relevant examples, including the diffraction by edges and the effects emerging from a minimal spacing. We established the mathematical form of the solutions and gave a detailed comparison with a problem in two dimensional space in a periodic background. A possible realization has been proposed in tight-binding arrays. The wide interest in photonic structures suggests applications of our results in this area, as well as solid state physics in time domain.

\ack

The author is pleased to thank Professor William Case for the many discussions on diffraction held during our stay in Ulm Universit\"at. The useful comments from an anonymous referee are also appreciated. Financial support from CONACyT Project 168752 {\it repatriaci\'on 2011,\ }is acknowledged. 

\appendix

\section*{Appendix A. Paths of a fixed length}

\setcounter{section}{1}
Here we find the number of paths of a constant lenght $S$. We are looking for all the solutions of the equation $S_{N+1,0}=S=\textrm{constant}$, subject to the constraints $\nu_0=m, \nu_{N+1}=n$. These conditions lead to configurations in which the segments of the paths can change directions, displacing from a point to its neighbours discontinuously. For each path we can define $S_l$ as the number of total steps going left and $S_r$ for steps taken to the right. We obviously have

\bea
S=S_l + S_r
\eea
Such steps can be taken in any order, but conspire to give the same total length $S\equiv S_{N+1,0}$ and the same initial and final position. Counting them is equivalent to find the number of combinations of $S_r$ segments in $S$ places (equivalently $S_l$ in $S$ places). As the partitions matter but not the permutations among fixed segments, we have the binomial distribution 

\bea
C(S)= \frac{S!}{S_r!S_l!}.
\label{ae25}
\eea
Now, $S_r$ turns out to be a fixed number, independent of the number of time slices $N$. It corresponds to the maximum displacement to the right in a path with a single change in direction (see the blue path in the left panel of figure \ref{fig:1}): $S_r = \frac{1}{2}(S+n-m)$ which is always an integer, as $S$ and $n-m$ must have the same parity. From this condition we also obtain $S_l$. Therefore we have

\bea
C(S)= \frac{S!}{(\frac{S+n-m}{2})!(\frac{S+m-n}{2})!}.
\label{ae26}
\eea
With this calculation, we can find the Feynman kernel as indicated in section 2.2.

\section*{Appendix B. Derivation of an operational formula}

Here we derive the approximation (\ref{e91}). Consider the general identity

\bea
f \left(\frac{d}{d x} \right) g(x)= g(x) f\left( \frac{d}{d x} + \frac{g'(x)}{g(x)}\right) [1]
\label{b1}
\eea
for $f$ and $g$ analytic and $[1]$ the constant function. This identity can be derived easily by noting that 

\bea
\frac{d g(x)}{d x} = g(x) \left( \frac{d }{d x}+\frac{g'(x)}{g(x)}\right)[1]
\eea
and similarly for its powers. For simplicity define now $h(x)= g'(x)/g(x)$. Taylor-expanding $f$ on the r.h.s. of (\ref{b1}) leads to 

\bea
f \left( \frac{d}{d x} + h(x)\right) = \sum_n \frac{ f_n(0)}{n!}  \left( \frac{d}{d x} + h(x) \right)^n.
\label{b2}
\eea
Now we expand the binomial $ \left(d /d x+ h(x)\right)^n$  and keep the only term without derivative operators (they act on the constant function $[1]$, therefore they vanish). The resulting expression contains terms of the form $ h(x)^a \partial_x^b h(x)^c$ with $a+b+c=n-1$. From these terms we only keep first derivatives, \ie $b=1$. This corresponds to small inverse powers of the energy $E$ in (\ref{e91}) whenever $x=iz\sqrt{E}$ and $g(x)=e^x$. This approximation leaves us with the terms

\bea
\sum_{j=0}^{n-2} h(x)^j \partial_x h(x)^{n-j-1} = \frac{n(n-1)}{2} h(x)^{n-2} h'(x) 
\label{b3}
\eea
as the only contribution for each term in the Taylor expansion. Substituting this back in (\ref{b2}) yields

\bea
 \sum_n \frac{ f_n(0)}{n!}  \frac{n(n-1)}{2} h(x)^{n-2} h'(x) = \frac{h'(x)}{2} f''(h(x)).
\label{b4}
\eea
Finally, we have

\bea
 f\left( \frac{d}{d x} + \frac{g'(x)}{g(x)}\right) [1] = f \left( \frac{g'(x)}{g(x)} \right) + \frac{1}{2} \left(\frac{g'(x)}{g(x)} \right)'  f'' \left( \frac{g'(x)}{g(x)} \right) + \dots
\label{b5}
\eea
With this expansion we can give certain limits for the operational form of our propagator (\ref{e89}).

\setcounter{section}{1}


\section*{References}

\begin{thebibliography}{99}

\bibitem{10} Albert J P \etal 2002 \emph{Optical and Quantum Electronics} {\bf 34} 251-263.

\bibitem{9} Heeger A J \etal 1988 \emph{Rev Mod Phys} {\bf 60} 781–850.

\bibitem{7} Yablanovitch E 1987 \emph{Phys Rev Lett} {\bf 58} 2059-2062.


\bibitem{hooft1} 'tHooft G 2012 Relating the quantum mechanics of discrete systems to standard canonical quantum mechanics. \emph{Preprint} quant-ph/1204.4926v1.

\bibitem{hooft2} 'tHooft G 2010 \emph{Int J Mod Phys} {\bf 25 } 4385-4396.

\bibitem{20} Sadurn\'i E, Seligman T H and Mortessagne F 2010  \emph{New J Phys} {\bf 12} 053014.

\bibitem{2} Moshinsky M 1952 \emph{Phys Rev} {\bf 88} 625-631.

\bibitem{bloch} Bloch F 1928 \emph{Z Phys} {\bf 52} 555-600.

\bibitem{1} Watson G N 1996 \emph{A Treatise on the Theory of Bessel Functions} (Cambridge University Press, reprint 2nd edition).

\bibitem{5} Sadurn\'i E 2012 Propagators in two-dimensional lattices. \emph{In preparation}.

\bibitem{1.1} Dattoli G, Cesarano C and Migliorati M 2003 \emph{Int J Math} {\bf 4} 239-246.

\bibitem{1.2} Dattoli G, Migliorati M and Srivastava H M 2005 \emph{Journal of Computational and Applied Mathematics} {\bf 173} 149-154.

\bibitem{3} Gradshteyn I S and Ryzhik I M 2007 \emph{Table of Integrals, Series and Products} (Academic Press, 7th edition).

\bibitem{3.1} Feynman R P and Hibbs A R 1965 \emph{Quantum Mechanics and Path Integrals} (McGraw-Hill, New York).

\bibitem{3.2} Grosche C and Steiner F 1998 \emph{Handbook of Feynman Path Integrals} (Springer).

\bibitem{3.3} Schulman L S 1996 \emph{Techniques and Applications of Path Integration} (John Wiley and Sons).

\bibitem{note1} There is a curious counterexample. Some relativistic propagators can be given as sums of paths in a ``checker-board" model of space-time (p. 34 \cite{3.1}). Such paths are continuous and their local velocity is equal in magnitude to $c$. This discrete formulation does not apply to our non-relativistic framework.

\bibitem{4} Sadurn\'i E 2012 \emph{J Phys: Conf Ser} {\bf 343} 012106.

\bibitem{11} Goldemberg J and Nussenzveig H M 1957 \emph{Rev Mex Fis} {\bf VI.3 }117-126.

\bibitem{nist} NIST 2012 {\it Handbook of Mathematical Functions,\ }Chapter 10.20. Online at http://dlmf.nist.gov/10.20

\bibitem{sym} Moshinsky M and Sadurn\'i E 2008 \emph{Rev Mex Fis} S {\bf 54 }(3) 92-98.

\bibitem{note2} It is possible to construct an expansion of the propagator in terms of associated Legendre functions of variable $z$ by using Bessel operators acting on $z^{\nu}$ (the symbolic form of Gegenbauer equation, p.140 in \cite{1}). Another expansion of fast convergence can be obtained by noting that $e^{iz\sqrt{E}}$ is the generator of Bessel polynomials of variable $z$ and parameter $E+1$.

%\bibitem{8} St\"ockmann H-J and Stein J 1990 \emph{Phys Rev Lett} {\bf 64} 2215-2218.

\end{thebibliography}
\end{document}